# A Parametric and Feasibility Study for Data Sampling of the Dynamic Mode Decomposition: Spectral Insights and Further Explorations

Cruz Y. Li[1] (李雨桐), Zengshun Chen[2*](陈增顺), Tim K.T. Tse[3**](谢锦添), Asiri Umenga Weerasuriya[4], Xuelin Zhang[5] (张雪琳), Yunfei Fu[6](付云飞), Xisheng Lin[7] (蔺习升)

[1,3,4,6,7] *Department of Civil and Environmental Engineering, The Hong Kong University of Science and Technology, Hong Kong SAR, China*

[2] *Department of Civil Engineering, Chongqing University, Chongqing, China*

[5] *School of Atmospheric Sciences, Sun Yat-sen University, Zhuhai, China.*

[1] yliht@connect.ust.hk; ORCID 0000-0002-9527-4674

[2] zchenba@connect.ust.hk; ORCID 0000-0001-5916-1165

[3] timkttse@ust.hk; ORCID 0000-0002-9678-1037

[4] asiriuw@connect.ust.hk; ORCID 0000-0001-8543-5449

[5] zhangxlin25@mail.sysu.edu.cn; ORCID 0000-0003-3941-4596

[6] yfuar@connect.ust.hk; ORCID 0000-0003-4225-081X

[7] xlinbl@connect.ust.hk; ORCID 0000-0002-1644-8796

[*] Co-first author with equal contribution.

[**] Corresponding author

All correspondence is directed to Dr Tim K.T. Tse.

# Abstract

This work continues the parametric investigation on the sampling nuances of the Dynamic Mode Decomposition (DMD) under the Koopman analysis. Through turbulent wakes, the investigation corroborated the generality of the universal convergence states for all DMD implementations. It discovered the implications of sampling range and resolution---the determinants of the spectral discretisation by discrete frequency bins and the highest resolved frequency, respectively. The work reaffirmed the necessity of the *Convergence* state for sampling independence, too. Results also suggested that the observables derived from the same flow may contain dynamically distinct information, thus altering the DMD output. The static pressure and vortex identification criteria are optimal variables for characterising structural response and fluid excitation. The pressure, velocity magnitude, and turbulence kinetic energy fields also suffice for general applications, but the Reynolds stresses and velocity components shall be avoided. Mean-subtraction is recommended for best approximations of the Koopman eigen tuples. Furthermore, the parametric investigation on truncation discovered some low-energy states that dictate a system's temporal integrity. The best practice for order reduction is to avoid truncation and employ dominant mode selection on a full-state subspace, though large-degree truncation supports fair data reconstruction with low computational cost. Finally, this work demonstrated the synthetic noise resulting from pre-decomposition interpolation. In unavoidable interpolations to increase the spatial dimension $n$, high-order schemes are recommended for better retention of the original dynamics. Finally, the observations herein, derived from inhomogeneous anisotropic turbulence, offer constructive references for DMD on fluid systems, if not also others beyond fluid mechanics.

# Declarations

The authors have no conflicts to disclose, including financial and non-financial interests and intellectual property.

# 1. Introduction

Every continuous fluid system contains an infinite number of infinitesimal fluid particles as individual degrees of freedom. The infinite-dimensionality is also compounded by the nonlinearity in the inter-particle interactions, making proper characterisations of fluid systems immensely difficult without order reduction and linearisation. On top of it, analytical renderings must also confront the unsolved Navier-Stokes, deeming the pursuit of solutions a climb of the Everest proportions. Fortunately, recent advancements in data science discovered the possibility of data-driven solutions to fluid problems [1]–[4]. To this end, the Koopman operator theory [5], [6] with the recently-developed data-driven interpretation [3], [7] emerges as a brilliant method to overcome the issues related to dimensionality and nonlinearity, while completely bypassing the Navier-Stokes in execution [7]–[9]. Many pioneering works have applied the Koopman analysis, or its algorithmic subordinate the Dynamic Mode Decomposition (DMD), to fluid systems with success [10]–[15]. Today, the DMD has become a popular technique for fluid analysis.

However, just like any new mathematical hatchlings, uncertainties of the DMD accompanies its vast potentials. Our previous work [16] has identified the knowledge gap in the sampling nuances of the technique and, through an extensive parametric study, offered some practical guidelines to reach sampling independence. On this solid foundation, the present work continues the exploration of uncharted waters. Specifically, it aims to consolidate the universal convergence states discovered in [16] by an additional parametric validation to prove their generality. It will also rationalise the generality by unveiling the spectral implications of the sampling range and resolution, before offering engineering suggestions for future endeavours.

Furthermore, the present work will parametrically assess the effects of input observables, aiming to identify the optimal DMD input for fluid analysis. Truncation, a frequently adopted method for order reduction, will also be systematically investigated to pinpoint its role in the approximation of the Koopman eigen tuples. On the other hand, our previous work [16] highlighted the importance of the $m<n$ condition for avoiding the *Divergence* state. Pre-decomposition interpolation is a common technique to artificially increase the spatial dimension $n$, thus preventing divergence. The present work will evaluate its impact on the DMD output to demonstrate the feasibility of interpolation. Finally, throughout the 16-month parametric project, we were driven by the persistent goal to provide useful references for future efforts. Therefore, we also aim to offer some candid, *a posteriori*, and application-oriented advice in the conclusive remarks of each parametrised quantity.

In composition, Section 2 provides a concise recapitulation of the test subject and previous findings, sparing readers the trouble of cumbersome cross-referencing. Section 3 formulates the methodology of the parametric investigation. Section 4 discusses the generality and spectral implications of the universal convergence states. Section 5 presents the investigations of the input variable, truncation, and interpolation. Section 6 summarises the main findings of the work.

## 2. A Brief Recapitulation

### *2.1 The Turbulent Flow*

We begin by briefly reviewing the test subject and previous findings. The main test subject of this serial effort is a subcritical free-shear flow over an infinitely long and stiff square prism. In the stationary state, free-shear flows are not only self-similar but also share many characteristics within the extended family (*e.g.,* jets and mixing layers). We selected a free-shear flow precisely for its commonality, aiming to promote the applicability of our findings. The prototypical flow-over-prism configuration, or the prism wake, is geometrically simplistic but phenomenologically sophisticated [17]–[21]. While the infinite span prevents the occurrence of uninvited end effects [22], [23], the prism stiffness simplifies the bi-directional feedback loop of fluid-structure interaction into the mono-direction case, in which the structure receives fluid excitation without altering the field by its own motion. The simplification is universally adopted by many studies on external flows, for example, those involving large-scale structures and civil applications [21], [22], [24]–[27]. In essence, this work aims to present the most fundamental yet sufficiently sophisticated test case to encourage intellectual resonance with the broadest readership.

Fluid-wise, we selected the subcritical range at the Reynolds Number *Re=$U_\infty D/v$=2.2×10$^4$*, where $U_\infty$ denotes the free-stream speed, *D* denotes the prism side length, and *v* denotes the kinematic viscosity of the fluid. The *Re* represents a wide neighbourhood of phenomenological similitude, in which the shear layer undergoes the turbulence transition II [18], [20]. It, too, ensures the incompressibility of air (*M<0.3*) for the divergence-free condition of the Navier-

Stokes equations. Furthermore, it permits reasonable expense for high-fidelity simulations and with reliable sources of validation [27]–[29].

The turbulent flow was simulated by the Large-Eddy Simulation with Near-Wall Resolution (LES-NWR) defined by Pope [30]. Numerical data was preferred over wind tunnel or field measurement because of its high-dimensionality and noise-free nature. The simulation adopted the DNS domain from Portela et al. [27], except a *4D* instead of *πD* spanwise length for computational ease in the Euclidean space (**Fig.** 1). Our LES-NWR solved the spatially filtered incompressible Navier-Stokes equations on an exclusively structured grid, while the dynamic-stress Smagorinsky model simulated the subgrid-scale dynamics. Inlet perturbation was also spared to avoid synthetic turbulence.

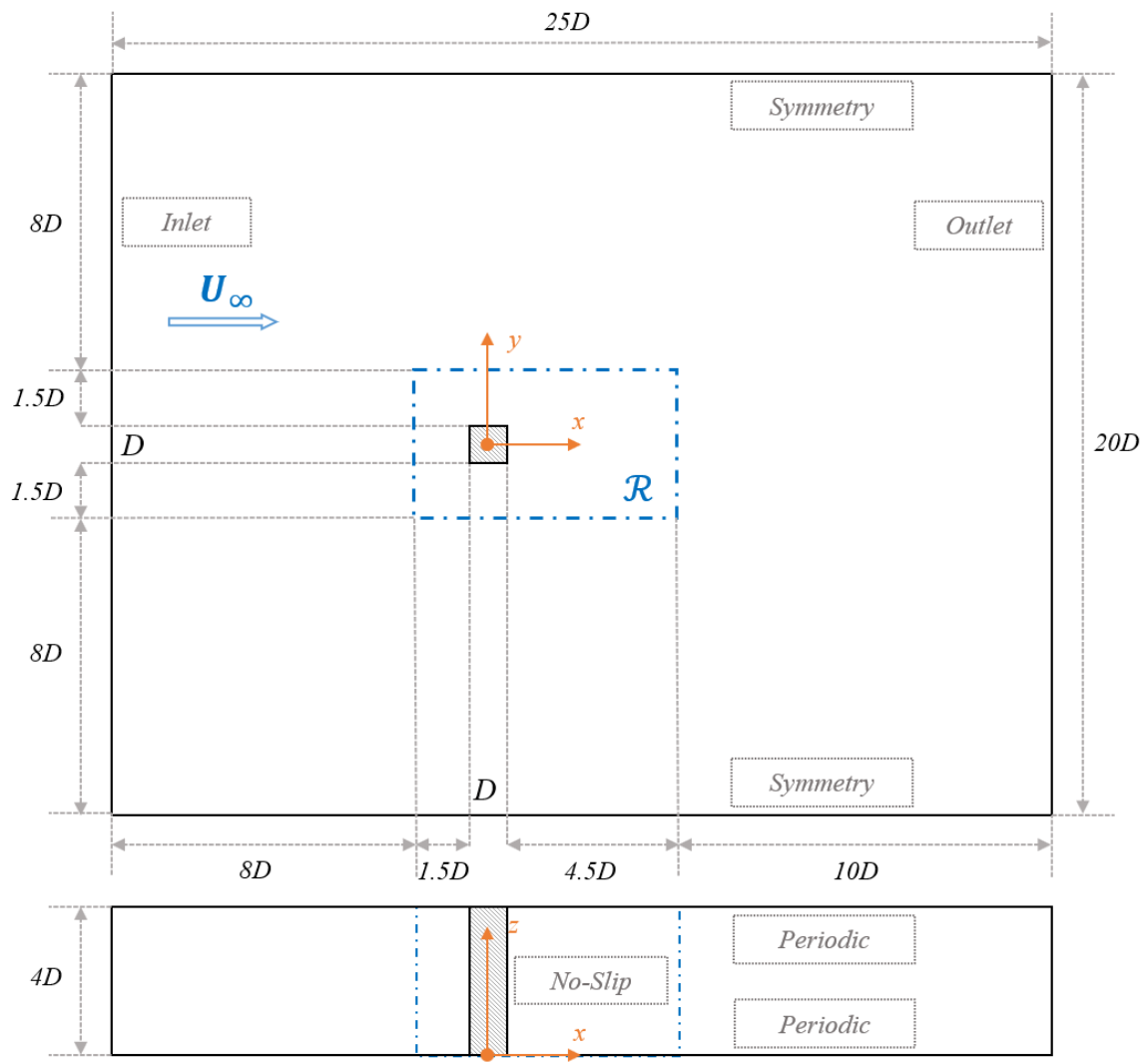

**Fig. 1** A schematic illustration of the computational domain and boundary conditions.

We deployed a finite-volume, segregated, pressure-based solution algorithm for this low-Mach-number incompressible flow. With second-order schemes for both spatial and temporal discretisation, and a stringent convergence criterion of $O^{-6}$, the effects of numerical dissipation and dispersion were minimized, respectively. Moreover, the simulation evolved by a non-dimensional time-interval $t^*$,

$$t^* = \frac{\Delta t \, U_\infty}{D} = 1.61 \times 10^{-3}, \qquad (3.1)$$

where $\Delta t$ is the physical time step, so the Courant-Friedrichs-Lewy (CFL) convergence condition is always satisfied, eliminating the possible time marching issues when solving partial differential equations.

The accuracy of the simulated flow is critical to the subsequent analysis. In short, the simulation resolves at least *80%* of the turbulent kinetic energy, qualifying itself as a proper LES-NWR. It also achieves an accuracy comparable to several DNS renderings [27]–[29]. To avoid repetition, we refer to our previous work [16] for full details about the simulation, grid resolution, and case validation. Appendix I also presents the essential information for the convenience of the readers.

## 2.2 Convergence of Sampling Range and Resolution

### 2.2.1 Universal Convergence States

The most significant finding of our previous investigation on the sampling range was four global convergence states: *Initialization, Transition, Stabilization*, and *Divergence*, as shown in **Fig.** 2.

The *Initialization* defines the early stage in which, by sampling only a small range of data, the DMD algorithm captures the spatiotemporal dynamics for fair reconstruction accuracy. However, it cannot find the optimal subspace nor a set of stable Ritz descriptors, so the individual modal characteristics are subject to great variability. This stage is characterised by fluctuating *St* and small reconstruction error.

The *Transition* marks the intermediate phase in which the algorithm seeks the optimal subspace with more sampled data. However, the trade-off is a drastic deterioration of reconstruction accuracy.

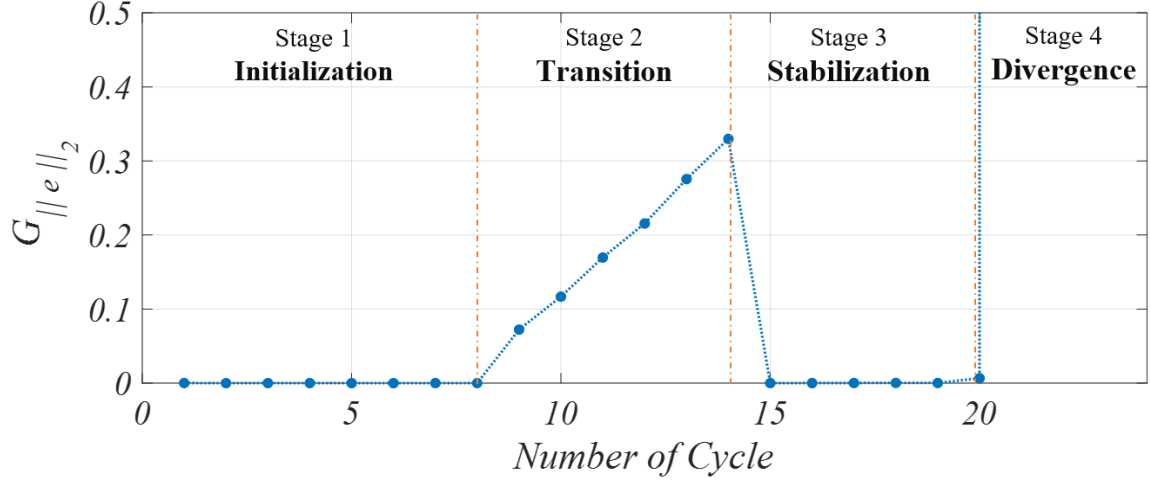

**Fig. 2** Grand mean $l_2$-norm of reconstruction error versus the number of DMD-sampled oscillation cycles of the prism wake.

The *Stabilization* describes the optimal state in which the algorithm fully establishes the convergence of modal characteristics and reconstruction accuracy. In this stage, sufficient data yields the optimal subspace, so the output of the decomposition becomes independent from sampling. This stage is characterised by near-constant *St* and minimal reconstruction error.

The *Divergence* depicts the case in which excessive sampling violates the DMD's tacit $m<n$ condition, so the Koopman operator loses fidelity, and the DMD algorithm yields degenerate output. This stage is characterised by the sudden loss of integrity in both the global and mode-specific DMD characteristics.

### 2.2.2 Bi-Parametrisation

Besides the convergence states, our previous bi-parametric investigation also revealed the combined effects of the sampling range and resolution, as summarised below and illustrated in **Fig.** 3:

- The convergence of the sampling range depends primarily on the global state of the system. The modal behaviours shift universally as the system transitions across the four states.

- The convergence of the sampling resolution depends primarily on the mode-specific periodicity. The convergence of one mode does not necessarily translate to that of another.
- For the fluid system herein, the convergence of the sampling range and resolution are mutually independent.
- For most analytical engineering implementations of the DMD, users are advised to sample a sufficient range to reach the *Stabilization* state without violating the $m<n$ condition and resolve target dynamics by at least 15 frames per cycle.
- For most reconstruction tasks, the *Initialization* state suffices. The *Transition* and *Instability* state shall generally be avoided.

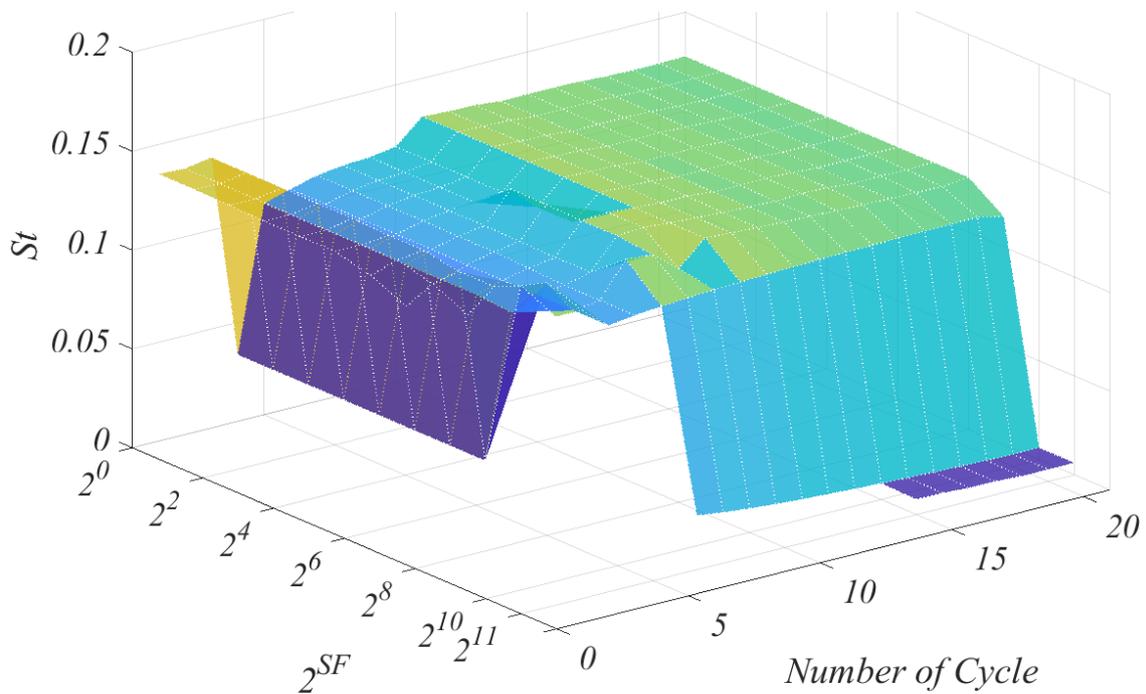

a)

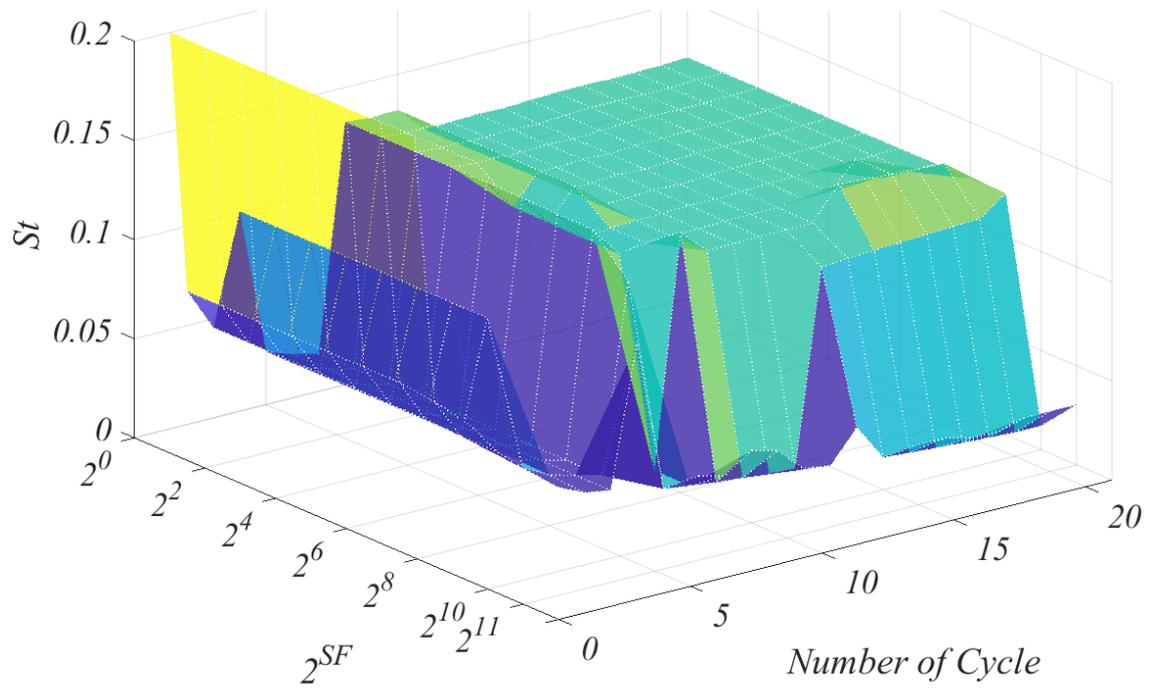

b)

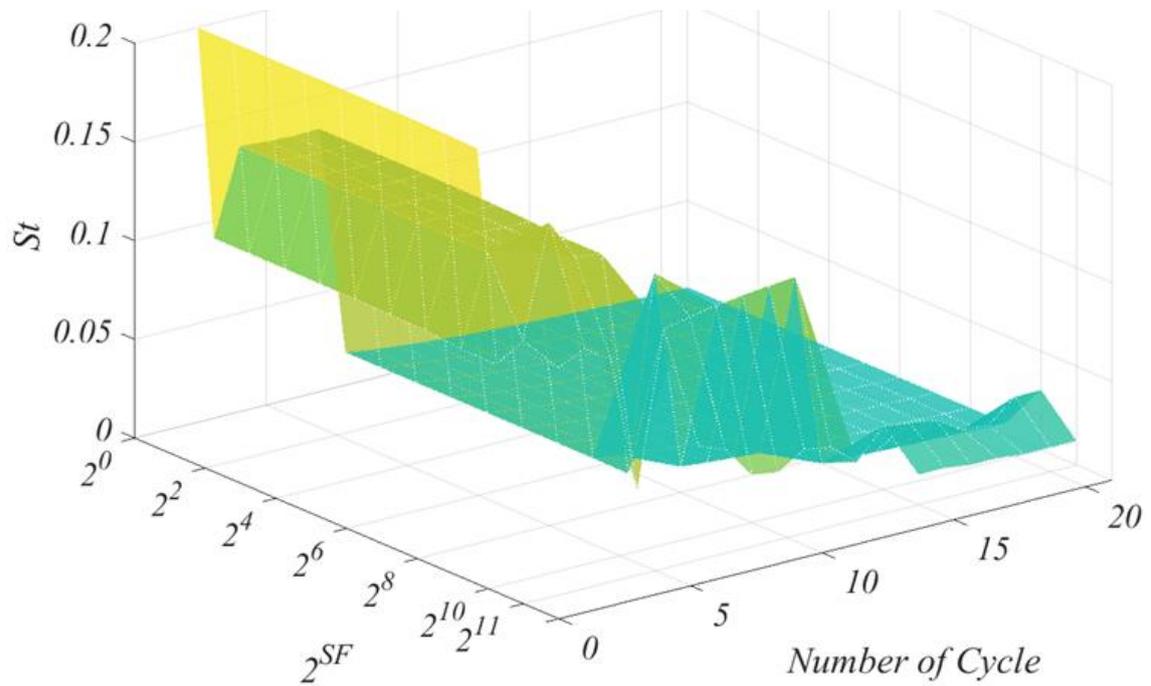

c)

**Fig. 3** The Strouhal number versus the sampling resolution $2^{SF}$ versus the number of DMD-sampled oscillation cycles of dominant a) Mode 1, b) Mode 2, and c) Mode 3.

# 3. Methodology

After the short reminder about the context and previous findings, the upcoming section will outline the methodology of the present investigation.

### *3.1 The Similarity-Expression Dynamic Mode Decomposition*

We begin by reminding the readers of the Koopman operator theory. According to Koopman [6], Mezić [3], and Rowley *et al*. [8], a dynamical system in discrete time can be expressed as

$$\bm{y}_{i+1} = \bm{f}(\bm{y}_i), \qquad (3.1.1)$$

where $i \in \mathbb{Z}$ and $\bm{f}$ is a map from a manifold $M$ to itself.

Then, one may impose the Koopman operator $U$, which is an infinite-dimensional linear operator that acts on scalar-valued functions on $M$, such that any scalar-valued function $g: M \to \mathbb{R}$, $U$ maps $g$ into a new function

$$U g(\bm{y}) = g(\bm{f}(\bm{y})). \qquad (3.1.2)$$

$U$ contains all the information of the dynamical system. If the system is linear, $U$ is exact. If the system is nonlinear, $U$ is a globally optimal, linear approximation. On this note, one may think of $U$ figuratively as a set of linear segments that discretise the nonlinear dynamics: given its infinite-dimensionality, the discretisation is infinitesimally fine, and the approximation is infinitely accurate.

The full-order Koopman operator exists only in theory. In practice, measuring any continuous system inevitably yields discrete data with finite spatial and temporal dimensions. It is precisely the inherent order-reduction that makes the practical realization of the Koopman operator theory possible. As pointed out by Williams et al. [31], several algorithms exist for computing the finite-dimensional Koopman approximation and its eigen tuples (*i.e.*, eigenfunctions, eigenvalues, Koopman modes), for example, the generalized Laplace analysis (GLA) [32]–[34] and the Ulam Galerkin method [35], [36].

For its algorithmic simplicity and robustness [31], [37], the Dynamic Mode Decomposition (DMD) [8], [10] is also one of the most used algorithms to approximate the eigenvalues and Koopman modes. The DMD has several subvariants, for example, the Arnoldi-based formulation, also known as the Koopman Mode Decomposition [8], the companion-matrix formulation [10], and similarity-matrix formulation [37]. The present work selects the similarity-matrix formulation because of its tractability with high-dimensional data [11]. The procedure for computing the so-called exact DMD modes is equivalent to an algorithm for finding the Koopman eigen tuples with linearly independent data snapshots [37].

Avoiding redundancy, we present only the quintessence of similarity-expression DMD here. Readers may refer to appendix II or Tu *et al.* [37] for the complete mathematics. The centrepiece of the DMD is the mapping matrix $A$ that connects two time-shifted sequences, $X_1$ and $X_2$. The sequences typically contain discrete snapshot data of a particular variable of interest, referred to as *observable*. $X_1$ and $X_2$ have the spatial dimension $n$ (rows) and temporal dimension $m$ (columns),

$$X_2 = AX_1. \qquad (3.1.3)$$

$A$ is an unknown matrix that mimics the map $f$, hence the Koopman operator $U$. Intuitively, its accuracy increases with dimensionality, but so too is the computational expense.

The similarity-expression DMD is a Singular-Value-Decomposition (SVD)-based algorithm that imposes a similarity-matrix $\tilde{A}$ to replace $A$ in **Eq.** 3.1.3. $\tilde{A}$ is the data-driven, reduced-order, and globally optimal approximation of $A$, hence the Koopman operator $U$. The deduction of $\tilde{A}$ is also exclusively implicit---it does not require any prior knowledge of a system's physics---making the DMD a promising route to bypass the unsolved Navier-Stokes.

After acquiring $\tilde{A}$ (see appendix II), one may easily arrive at its eigenvectors (Ritz vectors) $w_j$ and eigenvalues (Ritz values) $\lambda_j$ by the Ritz method, approximating the eigen tuples of the respective Koopman modes. To this end, the *exact* DMD modes [37] and the Koopman modes [8] are used interchangeably in this work, and can be expressed as

$$\boldsymbol{\Phi} = X_2 V \Sigma^{-1} W, \qquad (3.1.4)$$

where $\boldsymbol{\Phi}$ contains the DMD/Koopman mode $\phi_j$. $\Sigma$ and $V$ are outcomes of the SVD and contains the singular values $\sigma_j$ and temporally orthogonal modes $v_j$, respectively.

Every mode $\phi_j$ corresponds to a physical frequency $\omega_j$ in continuous time

$$\omega_j = \Im\{log(\lambda_j)\}/t^*, \qquad (3.1.5)$$

and a growth/decay rate $g_j$

$$g_j = \Re\{log(\lambda_j)\}/t^*, \qquad (3.1.6)$$

where $t^*$ is the uniform step of shift between $X_1$ and $X_2$.

One now acquires a Koopman system that constitutes a set of linear descriptors,

$$x_{Koopman,i} = \sum_{j=1}^{r} \phi_j \exp(\omega_j t_i)\alpha_j, \qquad (3.1.7)$$

which together offer the Koopman approximation of the input data at instant $i$. $r$ denotes the truncation order of $\tilde{A}$, and $\alpha_j$ denotes the coefficient of weight or the modal amplitude of $\phi_j$.

We quantify the accuracy of the Koopman approximation by the instantaneous, spatially-averaged, and $l_2$-normalized reconstruction error, $\|e\|_{2,i}$ or $\|e\|_{2,ins} \in \mathbb{R}^+$

$$\|e\|_{2,i} = \frac{X_{Koopman} - X}{X} = \frac{1}{n}\sum_{k=1}^{n}\left[\left(\frac{x_{Koopman,k,i} - x_{k,i}}{x_{k,i}}\right)^2\right]^{1/2}. \qquad (3.1.8)$$

To combine nodal and instantaneous error into a single spatiotemporal index, the grand mean $l_2$-norm of reconstruction error is defined as

$$G_{\|e\|_2} = \frac{1}{m}\sum_{i=1}^{m}\|e\|_{2,ins,i}. \qquad (3.1.9)$$

## *3.2 Inventory of Observables*

In total, 18 observables have been sampled as independent realizations of the fluid system, as summarised in **Table** 1. The observables include both primary and derived quantities of the prism response and the flow field. The input sequence containing each observable has also been sampled, curated, and post-processed independently without any algorithmic cross-communication. Since the DMD algorithm is physics-uninformed, prior connections between the 18 sequences is strictly nil.

The subsequent text refers to the upstream (AB), top (BC), downstream (CD), and bottom (DA) walls according to the orientation defined in **Fig.** 4. After Liu [38], we also refer to the vorticity-based vortex identification criterion, namely $|\omega|$, as the first-generation, the eigenvalue-based criteria, namely $q$ and $\lambda_2$, as the second-generation, and the ratio-based criteria, namely $\Omega$ and $\tilde{\Omega}_R$, as the third-generation.

Table 1 Summary of the inventory consisting of 18 observables

| Primary Observable | | Secondary Observable | |
| --- | --- | --- | --- |
| Wall | Flow Field | Turbulence | Vortex Identification |
| BC | $P$ | $\langle u'v' \rangle$ | $|\omega|$ |
| DA | $u$ | $\langle u'w' \rangle$ | $q$ |
| AB | $v$ | $\langle v'w' \rangle$ | $\lambda_2$ |
| CD | $w$ | $\langle k \rangle$ | $\Omega$ |
|    | $|U|$ |  | $\tilde{\Omega}_R$ |

### 3.2.1 Wall – Primary Observables

The static pressure $p$, or gauge pressure relative to the operating pressure ($1atm=101{,}325\ Pa$), on the top (BC), bottom (DA), upstream (AB), and downstream (CD) wall make up the four primary observables that describe the structural response.

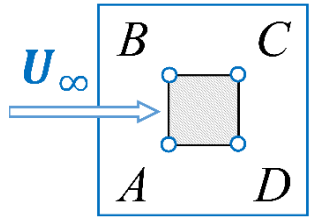

**Fig. 4** Schematic illustration of the prism walls.

### 3.2.2 Fluid – Primary Observables

The total pressure, or the pressure at the thermodynamic state that would exist if the fluid were brought to zero velocity and zero potential, was sampled for the flow field. For this specific low-Mach number flow, air is treated as incompressible, therefore the Bernoulli's equation,

$$P_o = p + p_{dyn}, \qquad (3.2.1)$$

is used to calculate the total pressure $P_o$, where $p$ is static pressure, $p_{dyn}$ is the local dynamic pressure

$$p_{dyn} = \frac{1}{2}\rho v^2. \qquad (3.2.2)$$

For simplicity, the wall static pressure and total pressure field will be denoted by $P$ in the ensuing discussions.

The velocity field generated four observables, namely the $u$, $v$, and $w$ components of the velocity vector $\boldsymbol{u}$ along the $x$-, $y$-, and $z$-axis in the Euclidean space, respectively, and the velocity magnitude

$$|U| = (u^2 + v^2 + w^2)^{1/2}. \qquad (3.2.3)$$

$P$, $u$, $v$, $w$, and $|U|$ make up the five primary observables that describe the flow field.

### 3.2.3  Fluid – Turbulence

Four turbulence-related observables were sampled for the flow field as well, including the Reynolds stress $\langle u'v' \rangle$, $\langle u'w' \rangle$, and $\langle v'w' \rangle$, as well as the turbulence kinetic energy $\langle k \rangle$

$$\langle k \rangle = \frac{1}{2}(\langle u'u' \rangle + \langle v'v' \rangle + \langle w'w' \rangle), \qquad (3.2.4)$$

where $u'$, $v'$, and $w'$, after the Reynolds decomposition, denote the fluctuating components of $u$, $v$, and $w$, respectively. Evidently, the turbulence observables appeal to the isotropic and deviatoric components of the Reynolds stress tensor.

### 3.2.4  Fluid – Vortex Identification Criteria

*First Generation: Vorticity-Based*

The last five observables are mathematically derived indices to identify vortical structures in a fluid flow. The first criterion is the most primitive, or the so-called first-generation criterion, vorticity magnitude $|\omega|$, developed after the vortex filament/tube concept of the Helmholtz's theorems [39]

$$|\omega| = (\omega_x^2 + \omega_y^2 + \omega_z^2)^{1/2} = |\nabla \times \boldsymbol{u}|. \qquad (3.2.5)$$

Despite the common utility of $|\omega|$, fundamental differences exist between vorticity and a vortex. Vorticity is the spin of an infinitesimal fluid particle and can be calculated for essential every point inside a fluid domain. A vortex, though its definition remains highly controversial till today, can be generally perceived as the rotation of a fluid region. A classic example is the laminar boundary layer, in which the viscous effect of the wall generates substantial vorticity but no rotational motion (i.e., vortex). On contrary to previous belief, the maximum vorticity magnitude also does not necessarily occur in the central region of a vortex [40].

*Second Generation: Eigenvalue-Based*

The second-generation criteria, as remarked by Liu *et al.* [40], are the eigenvalue-based indices that stem from the notion that a vortex can be seen as a region of closed or spiralling streamlines [41], [42]. These criteria are generally based on the velocity gradient, or the rate-of-strain, tensor, $L$,

$$\nabla u = \frac{\partial u}{\partial x} = L = \begin{bmatrix} \partial u/\partial x & \partial v/\partial x & \partial w/\partial x \\ \partial u/\partial y & \partial v/\partial y & \partial w/\partial y \\ \partial u/\partial z & \partial v/\partial z & \partial w/\partial z \end{bmatrix}. \tag{3.2.6}$$

Finding the three eigenvalues of $L$, $\lambda_1$, $\lambda_2$, and $\lambda_3$, the characteristic equation with Galilean invariance can be written as

$$\lambda^3 + p\lambda^2 + q\lambda + R = 0. \tag{3.2.7}$$

Galilean invariance means the result remains the same if the flow field undergoes a translation or superposition by a uniform field.

The three invariants of $L$ are

$$p = -tr(L) \tag{3.2.8}$$

$$q = -\frac{1}{2}[tr(\nabla u^2) - tr(L)^2] \tag{3.2.9}$$

$$r = -det(L) \tag{3.2.10}$$

For incompressible flows, the first invariant $p=0$.

The most common vortex identification criterion, the *q*-criterion, is proposed by Hunt et al. [43] directly according to the second invariant. The *q*-criterion is conveniently calculated and conceptually developed by the following expression,

$$q = \frac{1}{2}\left(\|\boldsymbol{Q}\|_F^2 - \|\boldsymbol{s}\|_F^2\right), \quad (3.2.11)$$

where $\boldsymbol{s}$ and $\boldsymbol{Q}$ are the symmetric and antisymmetric parts of $\boldsymbol{L}$,

$$\boldsymbol{s} = \frac{1}{2}\left(\nabla \boldsymbol{u} + \nabla \boldsymbol{u}^T\right), \quad (3.2.12)$$

$$\boldsymbol{Q} = \frac{1}{2}\left(\nabla \boldsymbol{u} - \nabla \boldsymbol{u}^T\right). \quad (3.2.13)$$

Theoretically, a region where $q>0$ and has a pressure lower than its surroundings defines a vortex. In practice, the pressure requirement is often overlooked, and a user-defined threshold must be defined to produce meaningful identification.

Another common criterion is the Galilean invariant $\lambda_2$-criterion proposed by Jeong & Hussain [44]. The basis for this criterion is the observation that a local pressure minimum in a plane fails to identify vortices under strong unsteady and viscous effects [40]. This method determines whether an arbitrary point in a flow field is a vortex core, thus finding the connected regions that make up a vortex. Formulation-wise, the $\lambda_2$ is calculated by finding the second eigenvalue of $\boldsymbol{Q}^2 + \boldsymbol{s}^2$, where $\lambda_1 \geq \lambda_2 \geq \lambda_3$. Theoretically, a vortex core is identified by $\lambda_2 < 0$. However, a user-defined threshold is also required in practice.

*Third Generation: Ratio-Based*

The primary limitation of the second-generation criterion is related to the user-defined threshold. As pointed out by Liu *et al.* [45], the threshold is case- and resolution-sensitive, so a proper definition is usually an esoteric and experience-based task. An improper definition may also result in issues like incorrect vortex capture, vortex breakdown, and the coexistence of weak and strong vortices. To provide a universal index, the original Ω-criterion was proposed [45].

The Ω-criterion is established upon the critical understanding that a vortex is a region where the vorticity overtakes deformation. Therefore, $0 \leq \Omega \in \mathbb{R}^+ \leq 1$ is defined as the ratio of the vorticity tensor to the sum of the vorticity and deformation tensors,

$$\Omega = \frac{\|\mathcal{Q}\|_F^2}{\|\mathcal{s}\|_F^2 + \|\mathcal{Q}\|_F^2 + \varepsilon_o}, \qquad (3.2.14)$$

where $\varepsilon_o$ is a small positive number to avoid division by zero. The threshold of the Ω-criterion is universal: *Ω>0.51* or *0.52* effectively detects a vortex.

On top of the Ω-criterion, an improved variant called the Omega-Liutex has been proposed recently [46]. The $\widetilde{\Omega}_R$-criterion adds the capability of defining the local rotational axis by considering the eigenvector of the velocity gradient tensor. Readers can refer to more details in [40], [47], [48]. The $\widetilde{\Omega}_R$-criterion $0 \leq \widetilde{\Omega}_R \in \mathbb{R}^+ \leq 1$ is defined as

$$\widetilde{\Omega}_R = \frac{\beta^2}{\beta^2 + \zeta^2 + \varepsilon_o}, \qquad (3.2.15)$$

where

$$\zeta = \frac{1}{2}((\partial v/\partial y - \partial u/\partial x)^2 + (\partial v/\partial x - \partial u/\partial y)^2)^{1/2}, \qquad (3.2.16)$$

$$\beta = \frac{1}{2}(\partial v/\partial x - \partial u/\partial y). \qquad (3.2.17)$$

## *3.3 Reynolds Decomposition, Truncation, and Interpolation*

### 3.3.1 Reynolds Decomposition for Mean-Subtraction

On top of the vast inventory of observables, the present work examines the DMD's sensitivity to the input data by comparatively assessing the fluctuating velocity ***u'***, instantaneous velocity ***u***, and mean velocity ⟨***u***⟩. For all three data sequences, 20 oscillation cycles are sampled, corresponding to the fully converged *Stabilization* state. The instantaneous, mean, and fluctuating velocities abide the Reynolds decomposition,

$$\boldsymbol{u} = \langle \boldsymbol{u} \rangle + \boldsymbol{u}' \qquad (3.3.1)$$

### 3.3.2 Truncation

This parametric study defines a truncation order $N_r$ to control the dimensionality $r$ of the SVD outcomes, hence that of the similarity matrix $\tilde{A} \in \mathbb{C}^{r \times r}$. The full-order sample mapped on the full-state subspace is chosen as 20 oscillation cycles of $u'$ in the *Stabilization* state. **Table** 2 summarizes the 23 truncation orders $N_r$ tested herein. The number of truncated modes $N_t = N_{r,full} - N_r$.

**Table 2** Summary of tested truncation order $N_r$ and corresponding number of truncated mode $N_t$.

| $N_r$ | $N_t$ |
|---|---|
| 19515 (Full-order) | 0 |
| 19514 | 1 |
| 19000 | 515 |
| 18000 - 1000 by increment of 1000 | |
| 500 | 19015 |
| 100 | 19415 |

**Pre-Decomposition Interpolation**

Data interpolation is a common if not indispensable step to visualising nonlinear systems. It also increases the spatial dimension $n$ of the input data, which may be a probable method to avoid divergence due to violating the $m<n$ condition. This work maps the original $u'$ field onto a non-uniform grid $G$ onto a newly generated equidistant grid $G_N$ to test the effect of interpolation. The spatial dimensions of $G_N$ and $G$ on the $x$-$y$ Cartesian plane is given by:

$$G_N = 3.33 G_x \cdot 3.33 G_y \sim 10 G \qquad (3.3.2)$$

We generated two interpolated datasets on $G_N$ using MATLAB's intrinsic *griddata* functions, namely the linear triangulation-based interpolation $f(u') \in C^0$ and cubic triangulation-based convolution $f(u') \in C^2$. Readers are directed to the MATLAB guide [49] for detailed mathematical formulations.

# 4. Spectral Insights into Convergence States

We now present the results and discussion of the present work. Though previous conclusions have unveiled the existence of the universal convergence states through the prism wake, we were bothered by a lingering, self-imposed scepticism: could the convergence states be just case-specific serendipity? Therefore, the first objective of the present work is to cast away the uncertainty, which motivated another parametric reinforcement with the cylinder wake.

*4.1 Generality of Universal Convergence States*

### 4.1.1   Prism-Cylinder Validation

The cylinder-prism cross-validation authenticates the generality of the convergence states by the following justifications. Fluid mechanically, the prism's rigidity forces direct separation and turbulent separation bubbles without boundary layers [50]. The leading-edge separation also encourages reattachment near the rear edge [21], [25]. By contrast, the smooth curvature of a cylinder induces boundary layers, a gradual separation, and laminar separation bubbles in sharp contrast to the prism case [51]. The delayed separation also lowers the chance for afterbody reattachment. The prism and cylinder wakes are geometrically similar for a straightforward comparison but phenomenologically distinct as two nonlinear systems of utterly different dynamics. Therefore, the prism-cylinder cross-validation will aptly testify for or against the generality of the convergence states.

All numerical details of the simulated cylinder wake are identical to those of the prism wake, except $D$ now denotes the diameter of the cylinder (**Fig.** 5). So, we refrain ourselves from repeating the same information. Another difference is that the curvilinear wall surface and the boundary layers tightened the grid requirement for the cylinder wake. After many iterative trials, a hexahedral grid of 8.0 million cells, almost twice the prism case, successfully resolved ⟨$k$⟩ and the $LES_{IQ}$ by at least 80%, attesting to a proper LES-NWR. All grid assessment and validation have been performed in the same fashion as the prism case in appendix I for a guarantee of input quality. Limited by concision, we will spare the presentation herein.

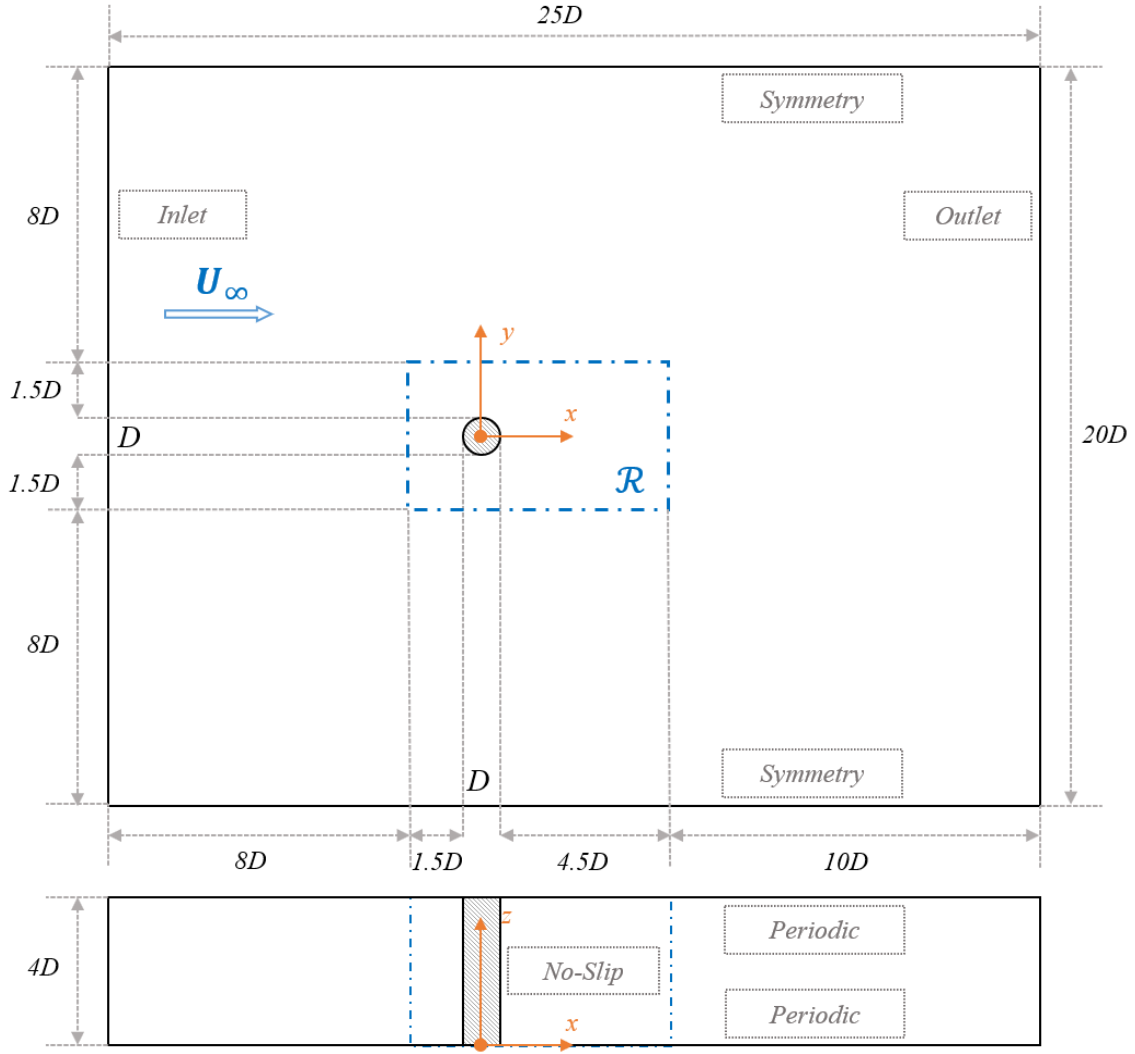

**Fig. 5** A schematic illustration of the computational domain and boundary conditions for the cylinder case.

### 4.1.2 Reconstruction Error

Analytically, we begin by examining the effect of the sampling range on the instantaneous reconstruction error, as presented in **Fig.** 6. By their behavioural disparity, cycles 1-18 are categorised as Group 1, cycles 19-50 as Group 2, and cycles 51-57 as Group 3. We emphasise the overall trend in the subsequent analysis, therefore skipping the extensive legend containing 57 entries. **Fig.** 6a illustrates that the instantaneous reconstruction error of Group 1 is overwhelming *<0.1* with only a few outliers *(<0.3)*. Group 1 exhibits excellent reconstruction accuracy throughout the sampled span, echoing the conclusion about the *Initialization* state from the prism case. Afterwards, Group 2 consists of a large span of cycles and displays the iconic error spike, marking the genesis of the *Transition* state (**Fig.** 6b). In this state, the error margin is universally elevated to about *0.5,* and the instantaneous error fluctuates thereabout.

The elevated error signals the intermediate phase in which the DMD sacrifices reconstruction accuracy to pursue an optimal subspace. Except for the more extended range of transition (31-50 cycles), the observations here are akin to the prism case. At last, the sudden restoration of the reconstruction accuracy occurs after sampling 51 cycles (**Fig.** 6c). Group 3 is representative of the *Stabilization* state. In this state, sufficient sampling ensures the DMD's establishment of the optimal POD subspace, so $\tilde{A}$ is temporally invariant. The DMD output also becomes pragmatically independent of the sampling range. Overall, the prism and cylinder wakes agree on the behaviours of instantaneous reconstruction.

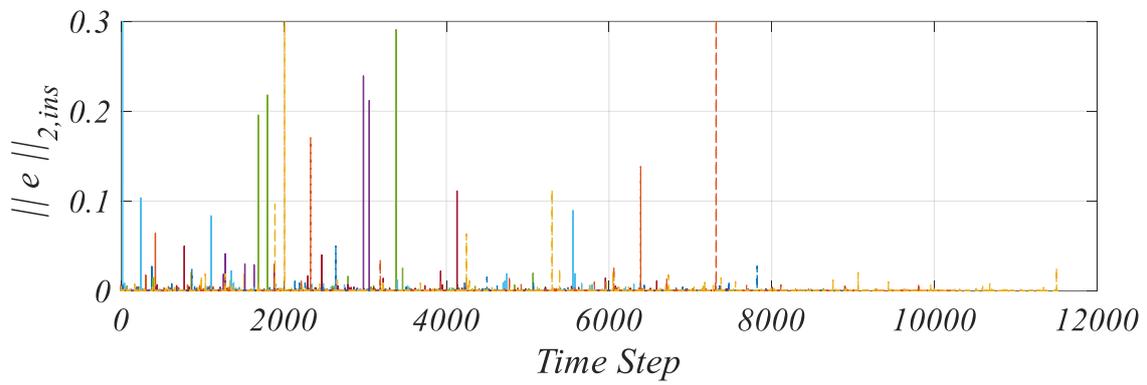

a)

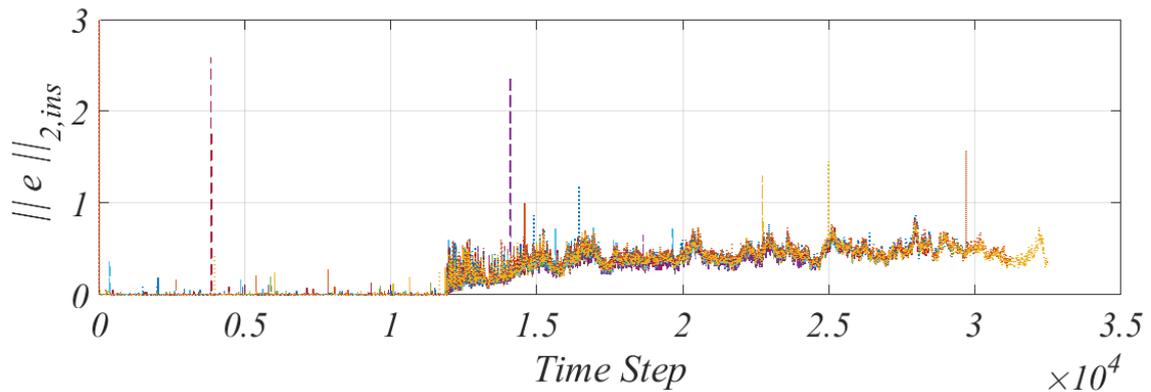

b)

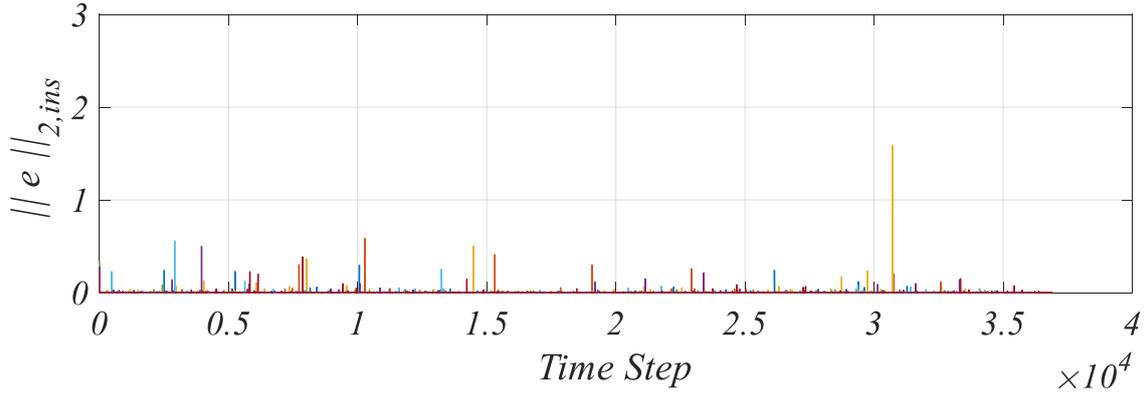

c)

**Fig. 6** Mean $l_2$-norm of reconstruction error versus time step $t^*$ of a) 1-18 (Group 1), b) 19-50 (Group 2), and c) 51-57 (Group 3) DMD-sampled oscillation cycles.

### 4.1.3 Reaffirming the Universal Convergence States

By no surprise, the grand mean reconstruction error (**Fig. 7**) of the cylinder wake identifies the *Initialization*, *Transition*, and *Stabilization* states, marking the initial reconstruction success, the subsequent transition, and the ultimate sampling convergence, respectively. Even the error build-ups in the *Transition* state are similar for the prism and cylinder wakes: the error margin displays a positive proportionality with the sampling range, before a sudden plummet stamps the bi-polar shift in system reconstruction after the transition into the *Stabilization* state. The near-identical trend in the grand mean error proves that the universal convergence states are not fortuitous nor case-specific, but a general feature of the DMD and the Koopman analysis.

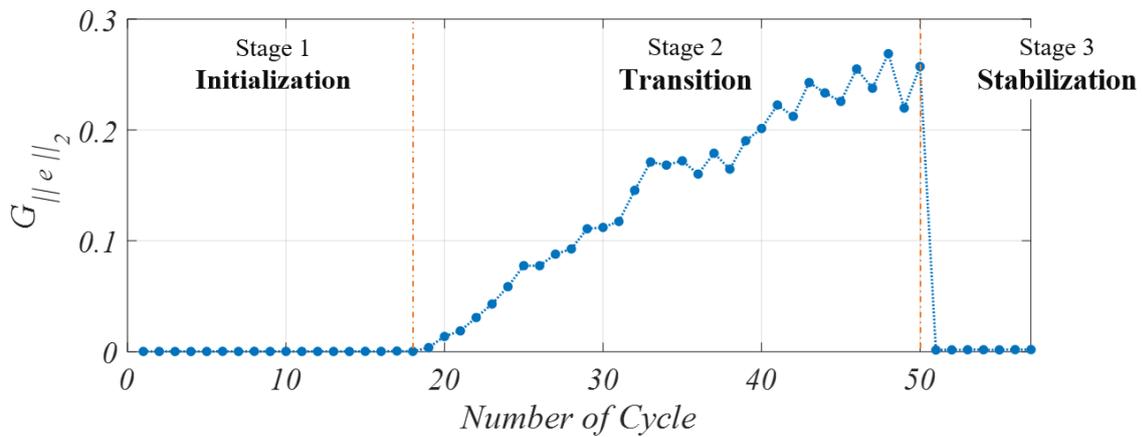

**Fig. 7** Grand mean $l_2$-norm of reconstruction error versus the number of DMD-sampled oscillation cycles of the cylinder wake.

A critical remark is that the parametric study of the cylinder wake did not sample until the violation of the $m<n$ condition so the *Divergence* state. The primary reason is the limitation of computational resources. To upkeep the integrity of the comparison, we maintained the same numerical and sampling timescales and data resolutions for both cases. However, the cylinder case required significantly more RAM and disc occupancy with nearly twice the grid size. In total, the serial effort already expended more than *800,000* core-hours for the DMD algorithmic implementation and more than 1.5 million core-hours with the LES-NWR. Considering the exponential growth of computational cost with the sample size, we stopped at 57 cycles after successfully underpinning the *Stabilization* state. Nevertheless, given the theoretical context of the DMD, one will certainly reach the *Divergence* state when sampling closely approaches or violates the $m<n$ condition.

### 4.1.4 The Bi-Parametric Validation

Finally, we validated the bi-parametric study of our previous work by the cylinder wake. **Fig. 8** presents the *St* of the dominant modes $M_{1-3}$ by considering the combined variations in sampling range and sampling resolution, as represented by the number of cycles and resolution index $2^{SF}$, respectively. For all three modes, the plateaus of *St* demonstrate the sampling convergence and the acquisition of a time-invariant $\tilde{A}$, agreeing with the prism wake.

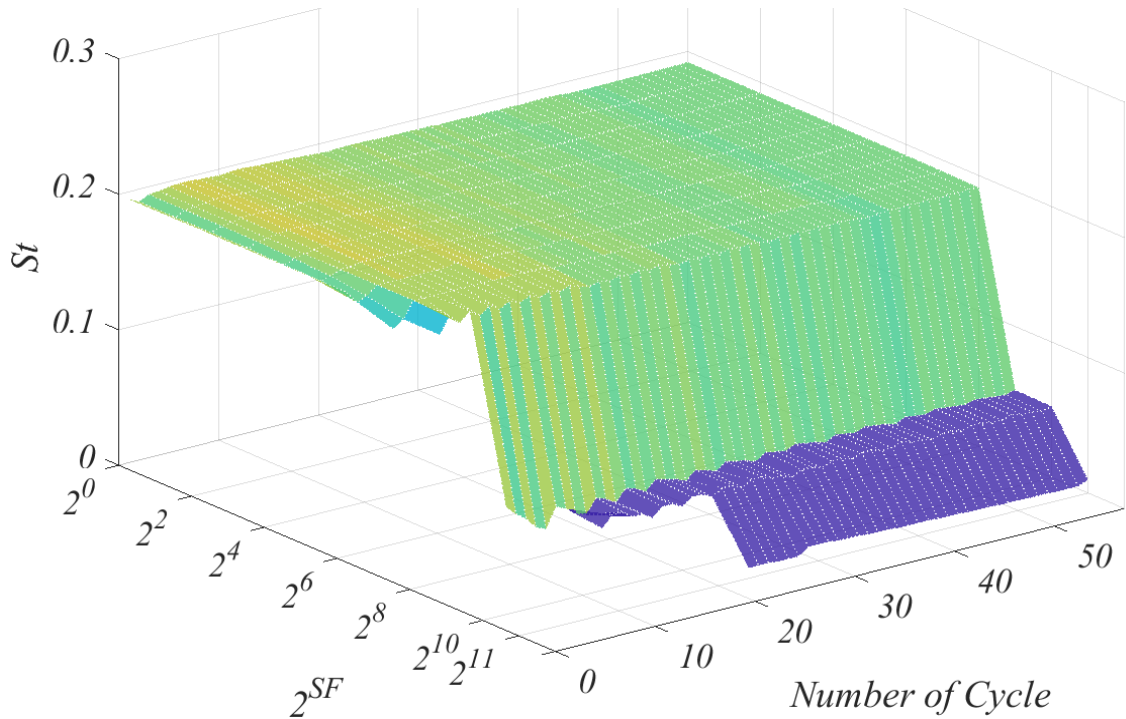

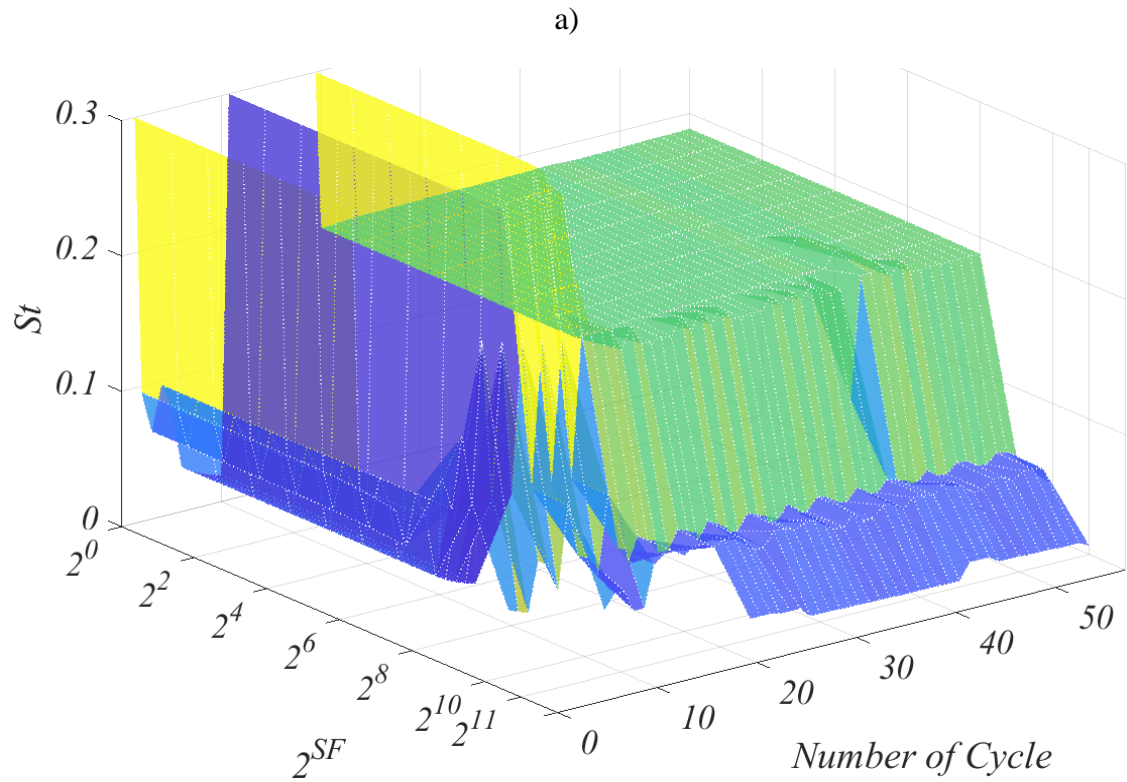

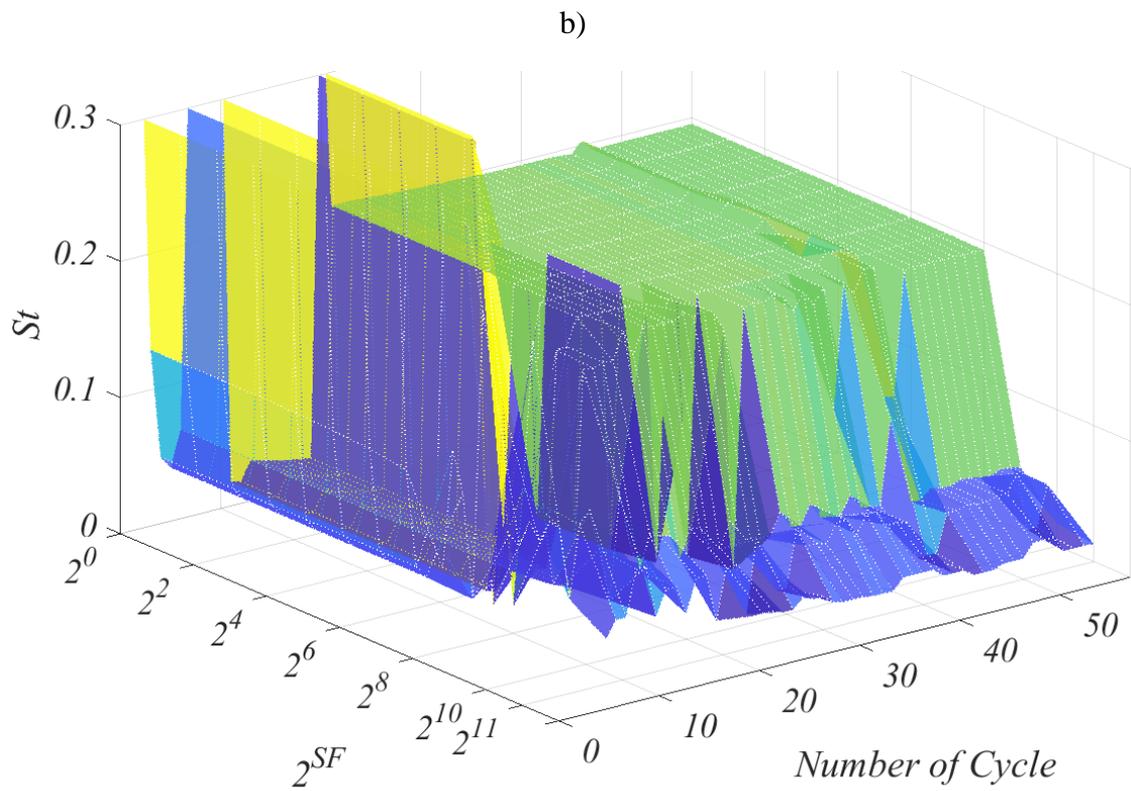

**Fig. 8** The Strouhal number versus the sampling resolution $2^{SF}$ versus the number of DMD-sampled oscillation cycles of dominant a) Mode 1, b) Mode 2, and c) Mode 3.

Moreover, the sampling convergence is only achieved when the range is *>19* cycles, and the resolution index is $2^{SF}<2^6$. Though an outlier is observed for $M_3$ with 46 cycles, the minute deterrence does not alter the overall trend. The implications, therefore, supports the notion that:

- The convergence of the sampling range depends primarily on the global state of the system, as it transitions across different convergence states.
- The convergence of the sampling resolution is mode-specific and dependent on modal periodicity.
- There is no apparent entanglement between the convergence of the sampling range and that of resolution.

Based on the consistent observations, we recommend users to:

1. Sample a sufficient range to reach the *Stabilization* state without violating the $m<n$ condition and resolve target dynamics by at least 15 frames per cycle for general engineering applications of the DMD.
2. Sample a small range to reach the *Initialization* state for reconstruction tasks with the DMD.
3. Avoid the *Transition* and *Divergence* states altogether.

## *4.2 Spectral Implications of Sampling Range and Resolution*

As previously concluded, the convergence of the sampling range is system-wise global, and that of the resolution is mode-specific. However, we could not help but to ponder on the question: what does the convergence imply and why are they achieved differently? For an explanation, we conducted an empirical investigation with three deliberately prescribed configurations, namely the *Coarse*, *Standard*, and *Fine,* as summarized in **Table** 3. The configurations are denoted by the subscripts $_C$, $_S$, and $_F$. All three configurations reached the *Stabilization* state. The *Coarse* and *Standard* differ only in the sampling range, while the *Standard*, and *Fine* differ only in the sampling resolution.

Ensuring generality, we tested the five observables for a total of 15 cases. The ten most dominant modes of BC, DA, AB, CD, and *u* are presented in **Fig.** 9 against *St* in the *Coarse*, *Standard*, and *Fine* configurations. The figure lucidly portrays the spectral implications of the sampling range and resolution. Comparing the *Coarse* and *Standard*, increasing the sampling range refines the discretisation of the spectrum as the frequency bins become increasingly narrow. The change is highlighted by the zoom-in image of $BC_C$ and $BC_S$ and agreed upon by all five observables. The bin width contracts from $St=0.0083$ to $0.0062$, but the highest resolved frequency is unchanged at $St=1.5557$. The observation implies that the frequency span over which the spatiotemporal content is averaged becomes smaller, hence sharper the bins and more accurate the discretisation. More importantly, once imposed, the contraction applies to every equidistant bin. Therefore, increasing the sampling range is equivalent to refining the spectral discretisation, and the *Stabilization* state marks the sufficiency in refinement. To this end, one shall expect the universal convergence states in essentially every DMD implementation, regardless of the nature of the input data.

**Table 3** Summary of the *Coarse*, *Standard*, and *Fine* configurations.

| Parameter | Configuration | | |
|---|---|---|---|
| | *Coarse* | *Standard* | *Fine* |
| Number of cycles | 15 | 20 | 20 |
| $SF$ ($2^{SF}$) | 6.32 (80) | 6.32 (80) | 8.32 (320) |
| Frames/cycle | 25 | 25 | 100 |

The empirical observation also gives rise to a fundamental insight. The *de jure* independence of the sampling range exists only in theory, if and only if the span of frequency bins becomes infinitesimal. Therefore, the convergence requirement of the DMD and the Koopman analysis shall not be absolute nor dogmatic. It must be pragmatic and decided upon the empirical observation that the frequency spectrum is sufficiently discretised. The criteria for the pragmatic convergence shall be bifold. First, the frequency discretisation shall consistently identify the dominant peaks. Second, if sufficiently significant, the bandwidth content of the peaks shall be adequately resolved. The Stabilization state adequately embodies the two requirements. For example, although microscopic differences are discernible between the *Coarse* and *Standard* configurations, their portrayals of energy concentration are macroscopically similar across the five observables, signalling achievement of the pragmatic sampling independence.

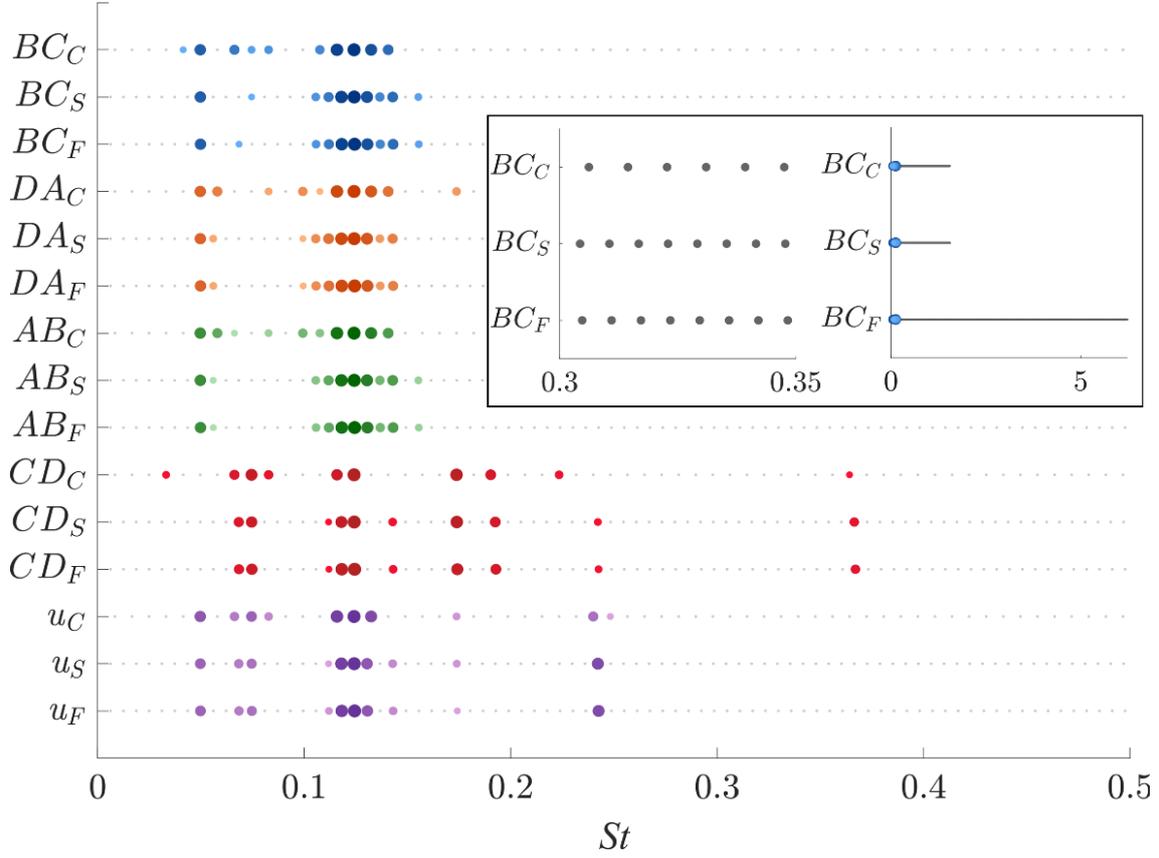

**Fig.9** Ten most dominant modes versus *St* of observables BC, DA, AB, CD, and *u* in the *Coarse*, *Standard*, and *Fine* configurations.

The spectral implications of the sampling resolution are illustrated by comparing the *Standard* and *Fine* configurations. Increasing the sampling resolution does not affect the discretisation but generates more equidistant bins towards the high-frequency spectrum. A comparison between $BC_S$ and $BC_F$ illustrates the unchanged bin width at *St=0.0062*, but the highest resolved frequency extends from *St=1.5557* to *6.2197*, precisely four times like the increase in the sampling resolution. The observation reinforces the previous conclusions about mode-specificity because alterations of sampling range add or erase spatiotemporal information altogether, affecting only to the added to erased modes *per se*. Therefore, the pragmatic convergence of sampling resolution only demands the inclusion of the most pertinent frequency range, so the dynamical activity of interest is effectively resolved.

The preceding analysis fully unveiled the spectral implications of the sampling range and resolution. The conclusions are paradoxical, fascinating, and almost mischievous. The sampling *range*, or the temporal dimension (number of snapshots) *m* of the input sequences, controls the *resolution* of the spectral discretisation, which globally affects every frequency bin. The sampling *resolution*, or the inter-snapshot time step of the input sequences, controls the *range* (upper limit) of the frequency spectrum, which projects an influence only to the affected bins. The messages from **Fig**. 9 also explain the generality of the universal convergence states. Regardless of the nature of the input data, the spectrum must have sufficient spectral resolution and contain the most relevant frequency range for the DMD to capture all the essential dynamics, hence achieving sampling independence in practice.

## 5. Further Explorations

*5.1 Input Observable*

### 5.1.1 Effects of Input

Besides the spectral implications, the vast inventory also facilitated the upcoming parametric investigation on the input variable. In the subsequent discussions, though analysing all $M_{1-20}$ and their respective Koopman eigen tuples, wewill only show the most representative $M_{1-3}$ for a concise presentation. As previously suggested, if $m$ of the input sequences is identical, the DMD generates discretisations of the same frequency bins $St_j$. The DMD spectra of all 18 observables in **Fig.** 10 confirms the notion, displaying poles that superimpose exactly onto one another. Moreover, the poles enclose the $\Re^2+\Im^2=1$ circle for a Region of Convergence (ROC) characteristic of acausal systems (mint green). Acausality means the system behaviours do not depend on past input but only on future ones [52], indicating the current sample captures all the major (or even minor) dynamical contributors. Again, the observation validates the sampling independence in practice.

On the other hand, the leading coefficient $\alpha_j$, as the index quantifying the dynamical significance of $St_j$, reflects the disparities in the spatiotemporal content of the input sequences. The need for inter-sequence comparison fostered the definition of the normalized modal amplitude,

$$-1 \leq |\tilde{\alpha}_j| \in \mathbb{R} \leq 1. \qquad (5.1.1)$$

Readers are reminded that $\alpha_j$ is the most rudimentary index of the DMD. Several other criteria [53]–[55] have also been developed, each with advantages and limitations. Till today, a universal consensus on the optimal index has yet to be reached, so the selection depends exclusively on user preference. Our scope here only concerns the most fundamental $\alpha_j$ index. Nevertheless, the same methodical procedure can be replicated for any given index. On this note, $|\tilde{\alpha}_1|$ of the most dominant mode is anchored at 1 for every observable in the upcoming analysis.

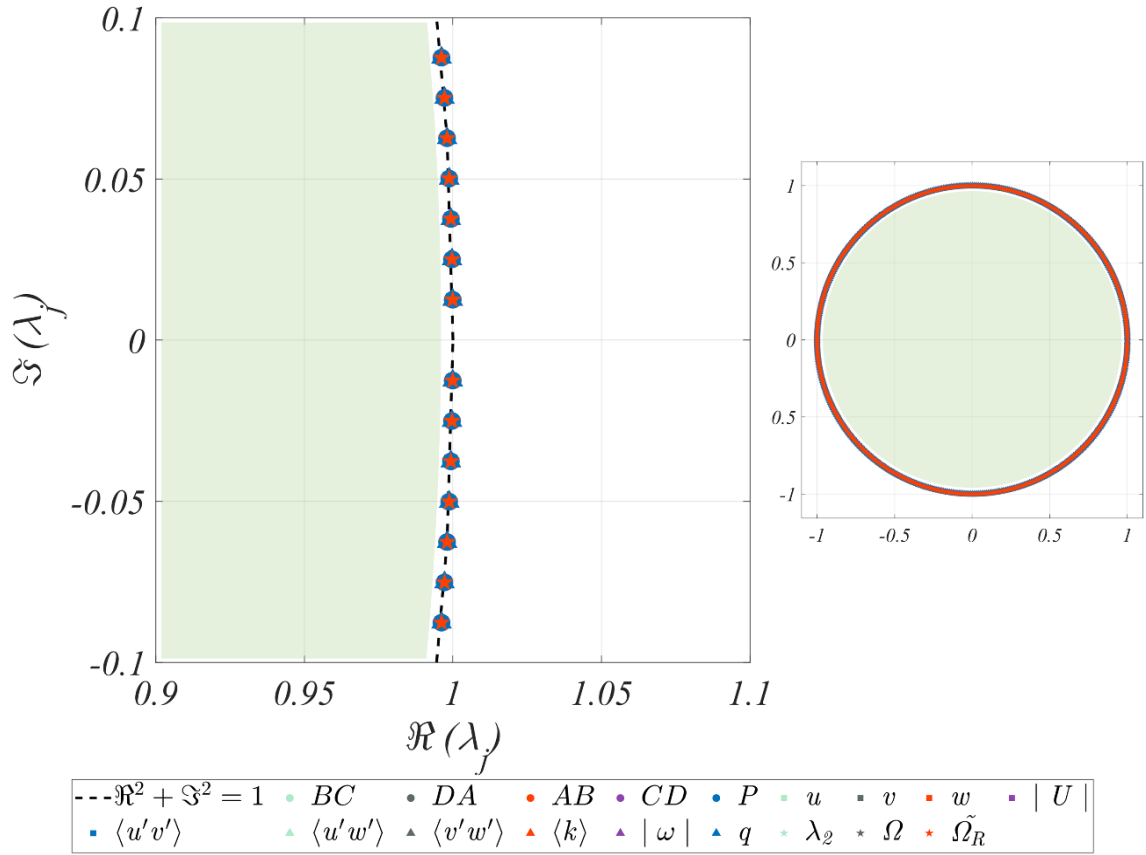

**Fig. 10** The DMD spectra showing the Region of Convergence of the 18 observables (left) and zoomed-in (right) near $\Re(\lambda_j)=1$ and $\Im(\lambda_j)=0$.

**Fig.** 11 presents the ten most dominant modes of the 18 observables versus *St*. Different classes of colour (blue, orange, green, and maroon) distinguish the four groups of observables, namely wall pressure, primary field variables, turbulence, and vortex identification criteria. The darkness of the colour and the radius of the markers figuratively illustrate the dominance of the Koopman modes. Accordingly, the location and concentration of the markers depict the energy concentration of the dynamics. Most straightforwardly, the input variable manifests profound impacts on the DMD output. Among the observables, the downstream wall (CD), *w* velocity, and the Reynolds stresses are markedly different from their peers, while others exhibit an apparent consensus.

The downstream wall (CD) displays the energy concentration of descending dominance at *St=0.1242, 0.1739, 0.0745, 0.1925, 0.3664,* and *0.2422,* while the other three walls, which we collectively refer to as the *on-wind* walls, show concentrations only at *St=0.1242* and *0.0497*. The two sets of prism walls are behaviourally distinct: the downstream wall reflects more interwoven and complex dynamics in the prism base, while the on-wind walls display more energetic potency in the broadband *St=0.1242* and the narrowband *St=0.0497* peaks. Fascinatingly, the flow field observables, except *w*, capture the dominant frequencies of the wall to a large degree. For example, *P* and *u* capture the peaks at *St=0.0497, 0.0745, 0.1242, 0.1739,* and *0.2422*. The vortex identification criterion captures all the wall dynamics, including the *St=0.1925* peak missed by *P* and *u*. Nevertheless, at least in terms of the top ten dominance, all field observables deem *St=0.3664* an insignificant peak. The analysis alludes to the existence of direct fluid-structure correspondences, which has been confirmed and systematically studied by our earlier works [56]–[58].

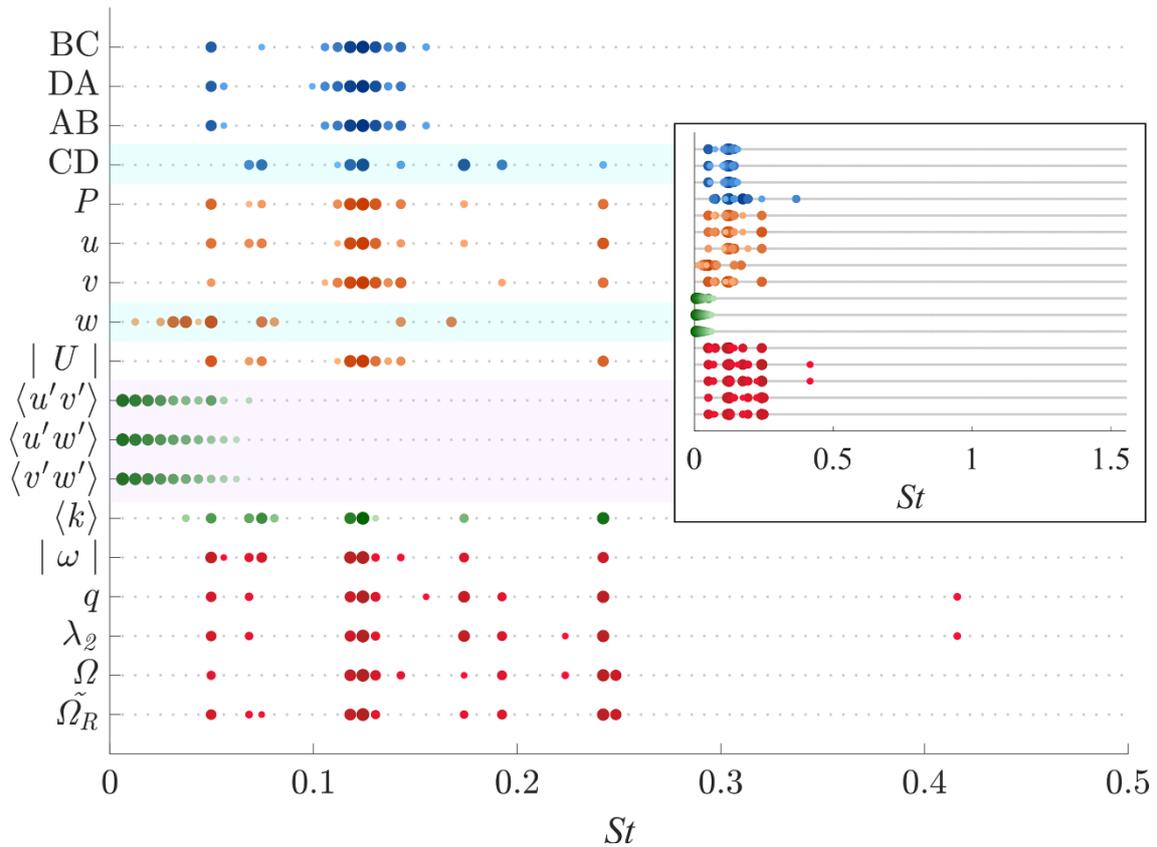

**Fig. 11** Ten modes with the highest $|\tilde{\alpha}_j|$ versus St of the 18 Koopman systems.

On the other hand, the singularity of *w* traces its origin to the predominance of streamwise convection in free-shear flows [30]. The energy distribution of |*U*| reflects the predominance as it is vividly overshadowed by the streamwise component *u*. On this note, the observation here buttresses the justifications of several previous efforts [59], [60], which had studied the three-dimensional prism wake by their planar counterpart. The DMD analysis also demonstrates the bleak influence of *w* in the subcritical prism wake.

The three components of the Reynolds stress are remarkably consistent in showing descending dominance from the lowest frequency. By definition, the deviatoric Reynolds stresses account for the turbulent fluctuations of fluid momentum and measure the distortion of infinitesimal fluid elements. Therefore, they quantify the tendency towards turbulence. As anticipated, the spatiotemporal content of the Reynolds stresses metaphorically depicts the energy cascade, in which the nonlinear inter-scale transfer can be both forward and inverse, but the overall energy balance is negative [27], [30]. The largest eddies, corresponding to the smallest wavenumbers or the lowest frequencies, are the most unstable and have the highest tendency towards turbulence, hence the greatest dominance.

Interestingly, the turbulence kinetic energy $\langle k \rangle$, calculated by the isotropic components of the stress tensor, is fundamentally different from its deviatoric cousins. $\langle k \rangle$ appeals to the distributions of the vortex identification criteria, implying that the dilatory fluid motions are highly related to vortical activities. Furthermore, except for $|\omega|$, the criteria capture the combined dynamics of all the prism walls, substantiating the notion that wall responses originate from the vortical activities. Otherwise, only minor dynamical differences are observed between the eigenvalue-based and the ratio-based criteria. Comparatively, $\langle k \rangle$ and $|\omega|$ are less than ideal for spectral characterisation.

In sum, the parametric study demonstrates that input observables, though derived from the same turbulent flow, may contain distinct dynamics, thus yielding vastly different DMD output. Consequently, selecting the most representative observables is critical. In fluid applications, while limited in the choice of structural response, one shall resort to the eigenvalue-based or ratio-based vortex identification criteria for the most indicative results. If post-processing is deemed mathematically cumbersome, the pressure, velocity magnitude, and turbulence kinetic energy fields generally suffice. The Reynolds stresses shall generally be avoided. The *u*, *v*, and *w* velocity fields can also be misleading unless the users possess fair understandings of the test subject. For example, the convection-predominant free-shear flow herein makes *u* a viable candidate but *w* a trivial representation.

### 5.1.2 Mean-Subtraction

Thus far, we have performed all the decompositions with mean-subtracted input. This section will parametrically examine the DMD's sensitivity to mean-subtraction through three highly correlated fields, namely the fluctuating velocity $u'$, instantaneous velocity $u$, and mean velocity $\langle u \rangle$. **Fig.** 12a shows marked differences in the absolute $St_{1-3}$ across the cases. However, $u'$ exhibit a similar trend as $u$ but differ drastically from $\langle u \rangle$. The dynamical disparity is reinforced by the growth/decay rate in **Fig.** 12b. The near-zero $g_{1-3}$ of $u'$ is in sharp contrast to $u$ and $\langle u \rangle$. For example, the absolute value of $g_1$ increases from $1.99 \times 10^{-7}$ of $u'$ to $1.87 \times 10^2$ of $u$, and $5.40 \times 10^1$ of $\langle u \rangle$, respectively. $g_2$ and $g_3$ also experience increases of at least five orders of magnitude. On this observation, the inclusion of the mean-field triggers acute deteriorations in the stability of the Koopman systems, reflecting the inadequacy of the Koopman eigen tuples in the portrayal of the mean-field dynamics.

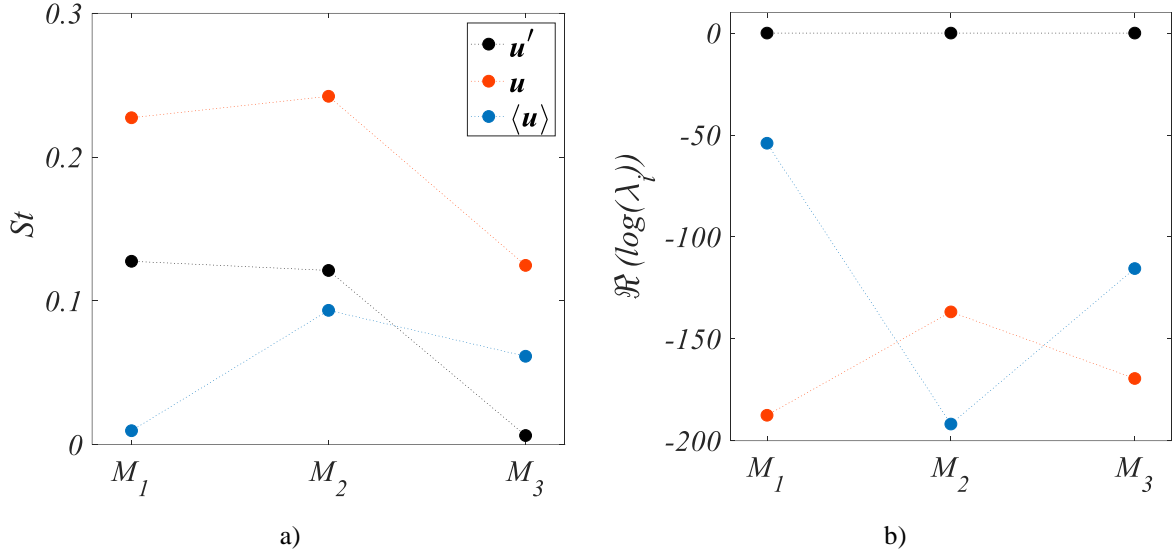

**Fig. 12** The a) Strouhal number $St_j$ and b) growth/decay rate $g_j$ of oscillatory DMD modes $M_{1-3}$ by sampling 20 oscillation cycles of fluctuating velocity $u'$, instantaneous velocity $u$, and mean velocity $\langle u \rangle$.

The DMD spectra paint the inadequacy in a spectacular portrait, as presented in **Fig.** 13. As expected, the poles of $u'$ are located infinitely close to the $\Re^2 + \Im^2 = 1$ circle, manifesting their practically perfect oscillations. The acausal, near-perennially stable system is contrasted by $u$ after including the mean-field. Poles inside the unit circle revert the system acausal to causal, terminating its independence of past input. The spectrum of the $\langle u \rangle$ *per se* further elucidates the aperiodicity of mean-field dynamics. The shrinking of the ROC into a semi-circle of only positive $\Re$ content meets the expectation that the mean-field is strictly real and appeals to no oscillatory motion. To help rationalise why *u'*, *u*, and $\langle u \rangle$ yield such substantial incongruences, one may think of the issue through the lens of turbulence, particularly the Richardson-Kolmogorov notion. Since *u'* symbolises turbulence, its decomposition into sinusoids and exponentials is analogous to representing eddies of different sizes by a wavenumber spectrum, which is intuitive. By the same analogy, the reluctant imposition of oscillatory descriptors herein is no different from enforcing the energy cascade onto the mean-field, which is bound to be erroneous.

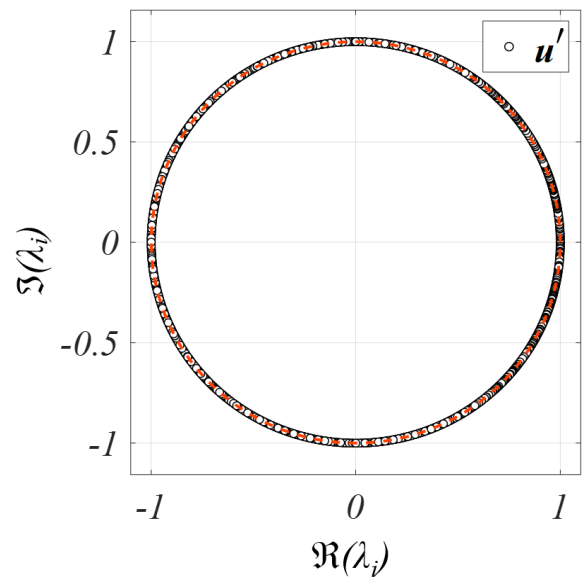

a)

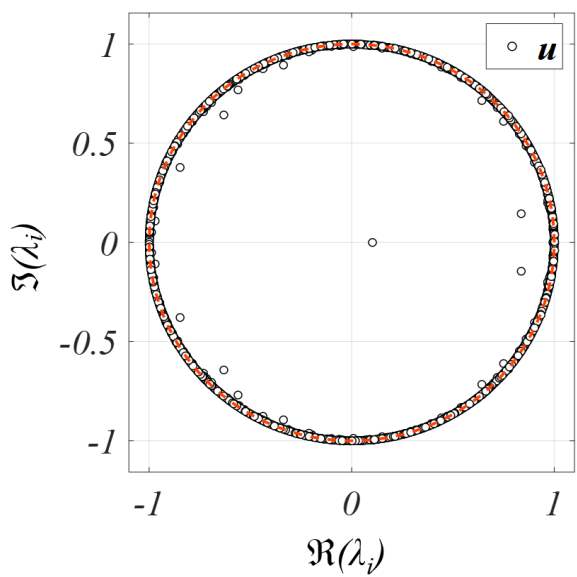 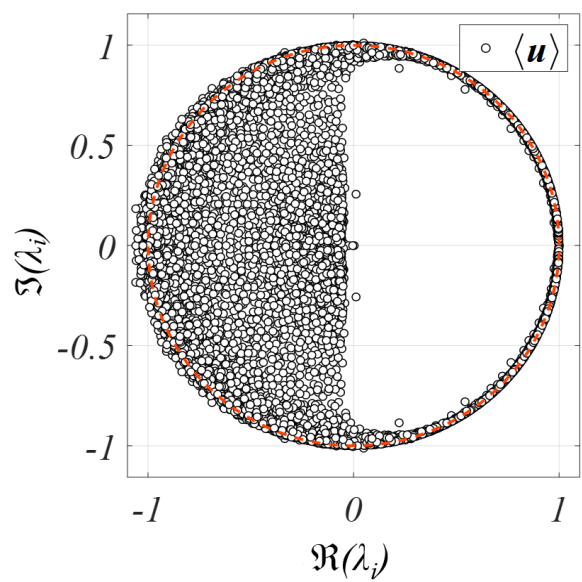

b)                                                                                          c)

**Fig. 13** The DMD spectra by sampling 20 oscillation cycles of a) fluctuating velocity $u'$, b) instantaneous velocity $u$ and c) mean velocity $\langle u \rangle$.

By no surprise, the duration of a valid Koopman description varies significantly across the three cases. **Fig.** 14 shows that the reconstruction error of $u'$ is consistently small with a few singularities towards the end of the sequence. With the mean-field, a reasonable reconstruction of $u$ only holds for about 2000 time-steps or so before shooting exponentially off to divergence. The DMD description of $\langle u \rangle$ is utterly invalid as the reconstruction error diverges immediately after the first few snapshots. The mode shapes of the most dominant Koopman mode $M_1$ for $u'$, $u$, and $\langle u \rangle$ have also been compared in Appendix III, confirming the different dynamical information carried by the leading eigen tuples.

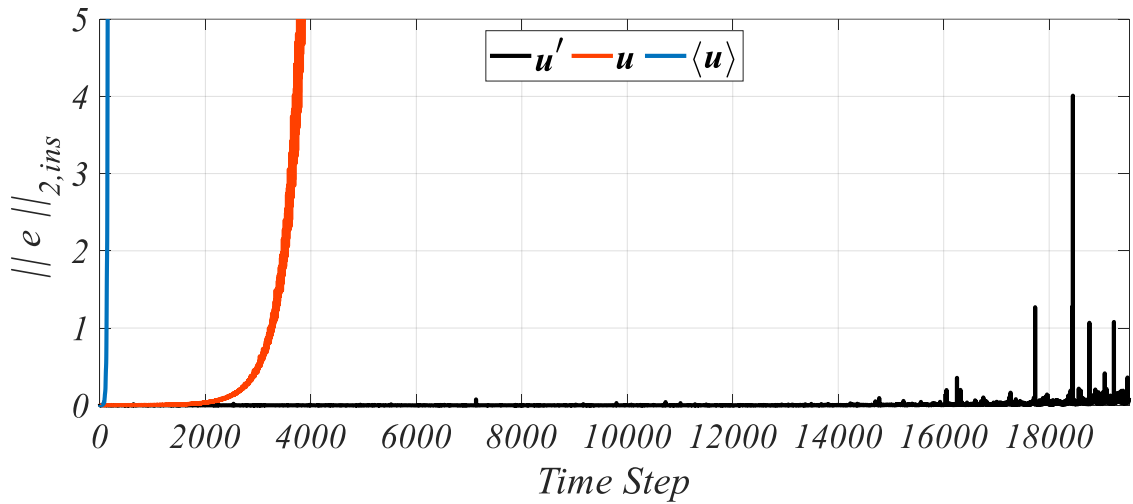

**Fig. 14** Mean $l_2$-norm of reconstruction error versus time step $t^*$ by sampling 20 oscillation cycles of fluctuating velocity $u'$, instantaneous velocity $u$, and mean velocity $\langle u \rangle$.

The origin of the DMD traces back to the Fourier transform, or the more general Z-transform. Mezić [3] has remarked that for any dynamical system with a Borel probability measure, the growth/decay rate is zero and Koopman modes are equivalent to Fourier modes. Stationary flows possess an ergodic measure by definition, so their Koopman modes are simply Fourier modes. Chen et al. [61] have also formally formulated the mathematical relationship between the DMD and the Discrete-Fourier transform (DFT) for zero-mean data taken from linearly independent snapshots. To this end, a DMD implementation with mean-subtracted data mimics the perennially oscillatory Koopman modes. One may also consider a mode's non-zero growth/decay rate as the artefact of the DMD approximation of the Koopman eigen tuples. Accordingly, the aforenoted inadequacy is in fact the undesired aftermath of forcing sinusoidal and exponential descriptions onto non-oscillatory motions. Therefore, although the DMD does not impose such a demand, we advise users always to deploy mean-subtracted data in practice unless the original data is already zero-mean. For fluid applications, the DMD characterisation is also more suitable for periodic flows, for example, those in the steady or stationary state.

## 5.2 Order-Reduction by Truncation

Switching gears, as a Reduced-Order Modelling (ROM) technique, the DMD reduces the dimension of an input system mainly in two ways. The first is dominant mode selection (d.m.s.), and the second is truncation.

### 5.2.1 Dominant Mode Selection vs. Truncation

We first clarify the different pathways the d.m.s. and truncation take to reduce a system's intrinsic order. The d.m.s. is a post-decomposition procedure that, based on available subspace, selects the most spatiotemporally relevant modes by some user-imposed criteria and threshold. For example, a selection based on $\alpha_j$ is typical of the d.m.s. Since it does not interfere with the order of $\tilde{A}$, the d.m.s does not compromise the computational expense of the algorithm. However, the direct benefit is that the integrity and stability $\tilde{A}$ remain intact. Our previous work showed that by the d.m.s., as little as *4%* of the data suffices to reconstruct the fluid system with less than *1%* reconstruction error [62].

On the other hand, truncation refers to the artificial curtailing of the POD subspace by manipulating the order of $\tilde{A}$. Expectedly, depending on $N_r$, the peri-decomposition procedure may marginally or significantly alleviate the computational expense. The analytical goal of this section is to evaluate the gains and losses as the result of truncation.

### 5.2.2 Singular Value and Low-Energy States

Typically, one determines $N_r$ according to the singular value $\sigma$, which measures the modal energy associated with $U$. **Fig**. 15 presents the singular value $\sigma$ versus truncation order $N_r$. As shown, all the tested $N_r$ (see **Table** 2) omit only less than *1%* of the total energy $\sum \sigma_i$. When $N_r$ *>500*, this margin is further reduced to *0.1%*. It is to say the truncated similarity matrix $\tilde{A} \in \mathbb{C}^{r \times r}$ considers all the major energy contributors for all the tested $N_r$ in this parametric study.

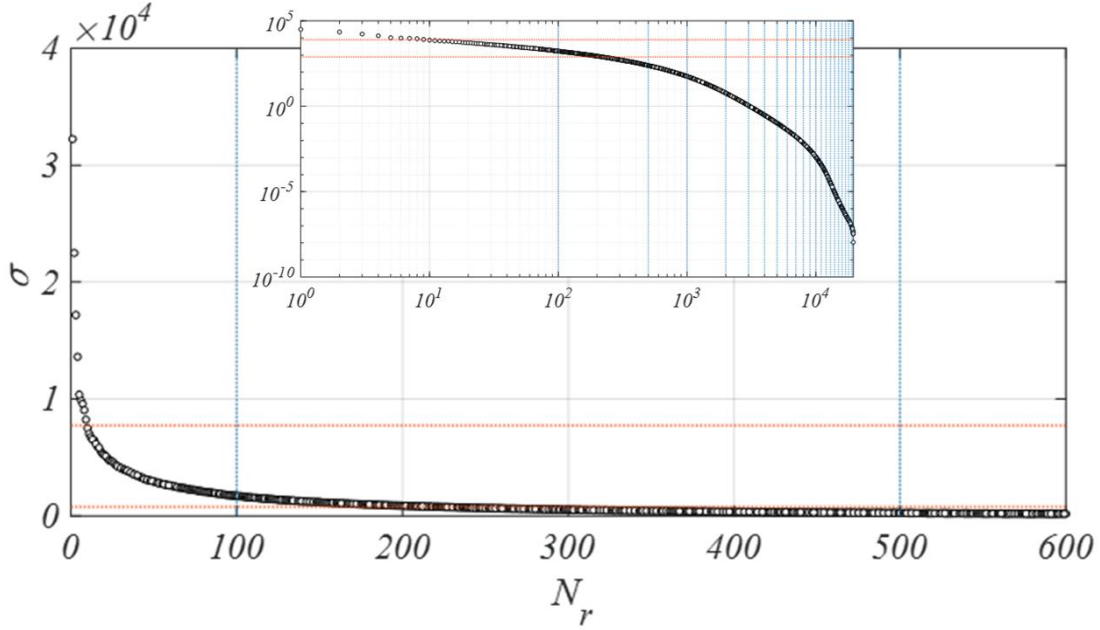

**Fig. 15** Singular value $\sigma$ versus truncation order $N_r$; Orange dash ---- *1%* and *0.1%* levels of total energy $\sum \sigma_i$; Blue dash ---- selected $N_r$. Inset figure in logarithmic scale.

We begin by considering $St_{1\text{-}3}$ versus the truncation order $N_r$, as presented in **Fig.** 16. $St_{1\text{-}3}$ in parenthesis are the *St's* of the full-order system, presented as the benchmark for the comparison. Clearly, truncation induces extreme variations in the reduced frequency, as changes can be as small as a few per cent to as large as three orders of magnitude. At first sight, the variations do not appeal to any apparent pattern and perhaps are even against intuition. For example, when $N_r=19514$, the omission of a single order causes the frequency of the leading mode $M_1$ to increase by nearly 200 times. However, minimal variations are observed in $St_{1\text{-}3}$ when $N_r=5000$ produces a purportedly much more significant truncation. No immediate rationale can be associated with the primitive observations. So, we must question whether the variations result from an absolute change in the spectral discretisation or are simply an alternation in dominance ranking.

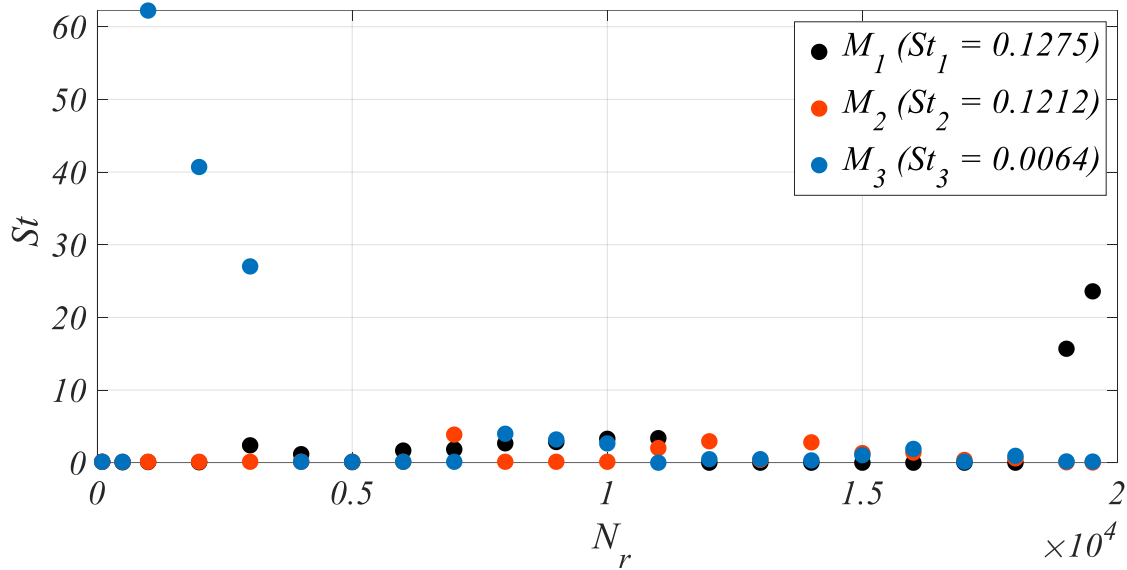

**Fig. 16** The Strouhal number *St* of DMD modes $M_{1-3}$ versus truncation order $N_r$; $St_{1-3}$: *St* of the full-order $M_{1-3}$.

Close inspections revealed both scenarios take place. **Table** 4 presents the sorted results of representative $N_r$ from **Fig.** 16. $N_r = 10,000$ stands out as a clear behavioural demarcation. When $N_r>10,000$, the variations are primarily attributed to altered dominance ranking, and the changes in the spectral discretisation are marginal. The dominance hierarchy between $M_{1-3}$ also remains unchanged, in a sense that $M_1$ is always more dominant than $M_2$ than $M_3$. In this case, despite the ranking differences, the nonlinear system is generally decomposed into a consistent set of Ritz descriptors to represent its dynamical behaviours.

By contrast, when $N_r<10,000$, $M_{1-3}$ become unrecognizable. We clarify that being unrecognisable yields a few possibilities. The most direct and likely possibility is the utter change of spectral discretisation, which signals a fundamental shift in the Ritz representation of the nonlinear system. Another possibility is that, for the same Ritz descriptors, the modal stability deteriorates so substantially to the extent that the modes are no longer identifiable. Finally, it is also possible that *M1-20* never changed, but they become so spatiotemporally irrelevant by dropping significantly in the dominance ranking. The upcoming reconstruction analysis will demonstrate that the cause is related to the spectral discretisation. Nevertheless, for now, and for all three possibilities, it is indisputable that the dynamical content of the truncated system differs dramatically from that of the original.

**Table 4** Summary of the Strouhal number *St* and respective $α_j$ ranking of DMD modes $M_{1-3}$ at representative truncation order $N_r$.

| $N_r$ | $M_1$ | $M_2$ | $M_3$ $(10^{-3})$ |
|:---:|:---:|:---:|:---:|
| | *St - [ranking]** | | |
| *Full-order* | *0.128 - [1]* | *0.121 - [2]* | *6.38 - [3]* |
| 19514 | *0.128 - [2]* | *0.121 - [3]* | *6.38 - [4]* |
| 18000 | *0.128 - [3]* | *0.121 - [5]* | *6.38 - [6]* |
| 15000 | *0.128 - [2]* | *0.121 - [3]* | *6.39 - [4]* |
| 10000 | *0.127 - [2]* | *0.121 - [3]* | *6.65 - [7]* |
| 9000 | *0.127 - [2]* | *0.121 - [4]* | *u.r.* |
| 5000 | *u.r.* | *u.r.* | *u.r.* |
| 1000 | *u.r.* | *u.r.* | *u.r.* |
| 500 | *u.r.* | *u.r.* | *u.r.* |

*u.r. – unrecognizable*

Based on the sorted result, **Fig.** 17 illustrates the behavioural demarcation as $M_3$ becomes unrecognizable. The conclusion is lucid: excessive truncation leads to substantial losses in the temporal integrity. At this point, one may also ponder the cause of the deterioration. Compared to *4%* of the d.m.s., why does truncation require nearly *50%* of the original data for satisfactory dynamical outputs?

We remind the readers about the selection criterion of truncation, the singular value *σ*. The SVD separates the input data's spatial and temporal components, and *σ* is strictly a spatial measure. However, as Schmid [10] pointed out and Noack *et al.* [63] empirically observed, some low-energy or even zero-energy states may have significant dynamical effects on the nonlinear system without contributing to the low-order POD modes. As observed in this work with $N_r = 9,000$, the loss of temporal integrity takes place even when the truncated subspace captures more than *99.9999999%* of the total energy. The result underpins the existence of the dynamically critical but low-energy states—the ones that constitute less than $10^{-8}$ *%* of the total energy but dictate the temporal integrity of the system. The relatively inferior performance of truncation compared to the d.m.s. also traces back to the neglect of these low-energy states. On this note, the existence of the dynamical low-energy states exemplifies a scenario where the DMD becomes an advantageous alternative to the POD for reduced-order modelling.

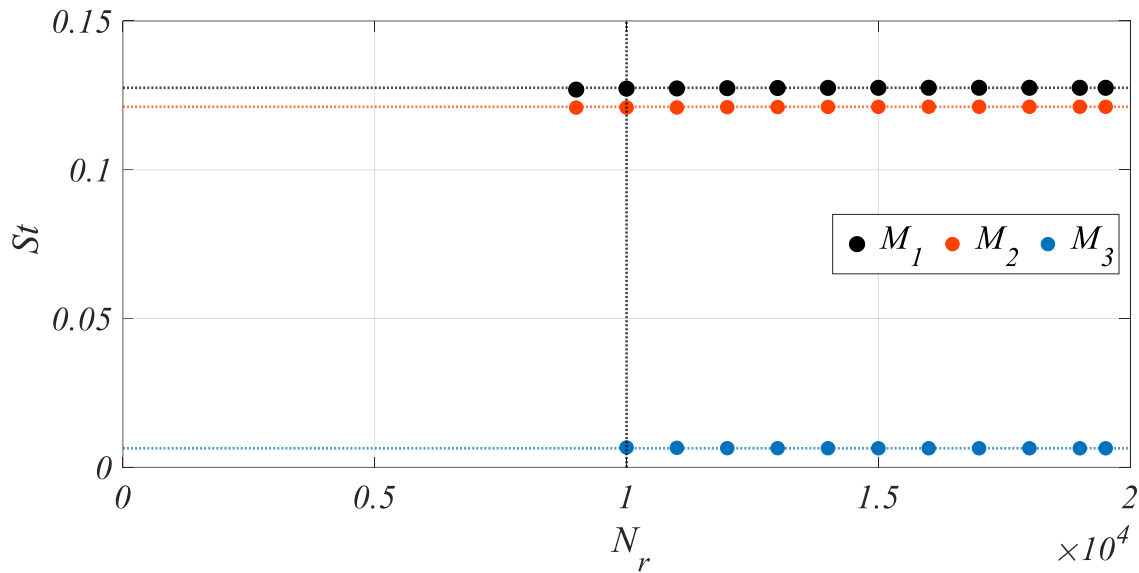

**Fig. 17** The Strouhal number *St* of DMD modes $M_{1-3}$ versus truncation order $N_r$ after mode sorting; Black dash ---- $St_1 = 0.1275$; Orange dash ---- $St_2 = 0.1212$; Blue dash ---- $St_3 = 0.0064$.

### 5.2.3 POD Subspace and Reconstruction

In addition to analysing the temporal integrity, the DMD spectra also figuratively illustrate how truncation affects the POD subspace and system stability, as shown in **Fig.** 18. The full-order benchmark exhibits stellar stability and acausality, attesting to the established sampling independence. Interestingly, by truncating only a single order ($N_r = 19514$), the stability and acausality experience notable deteriorations (**Fig**. 18b). Two diverging poles appear far outside the unit circle, while other poles appear inside to encroach the ROC. Truncation by as little as one dimension introduces instability and shifts the causality of a Koopman system. As $N_r$ further decreases, the deterioration becomes more significant, as poles on the -$\Re$ side increasingly intrude on the ROC, visually illustrating the shrinkage of the POD subspace. Since shrinkage begins from the high-frequency end (**Figs.** 18c-d), the low-order modes, representing the most dominant mechanisms, only sense the effect when truncation becomes sufficiently substantial (**Figs.** 18e-f).

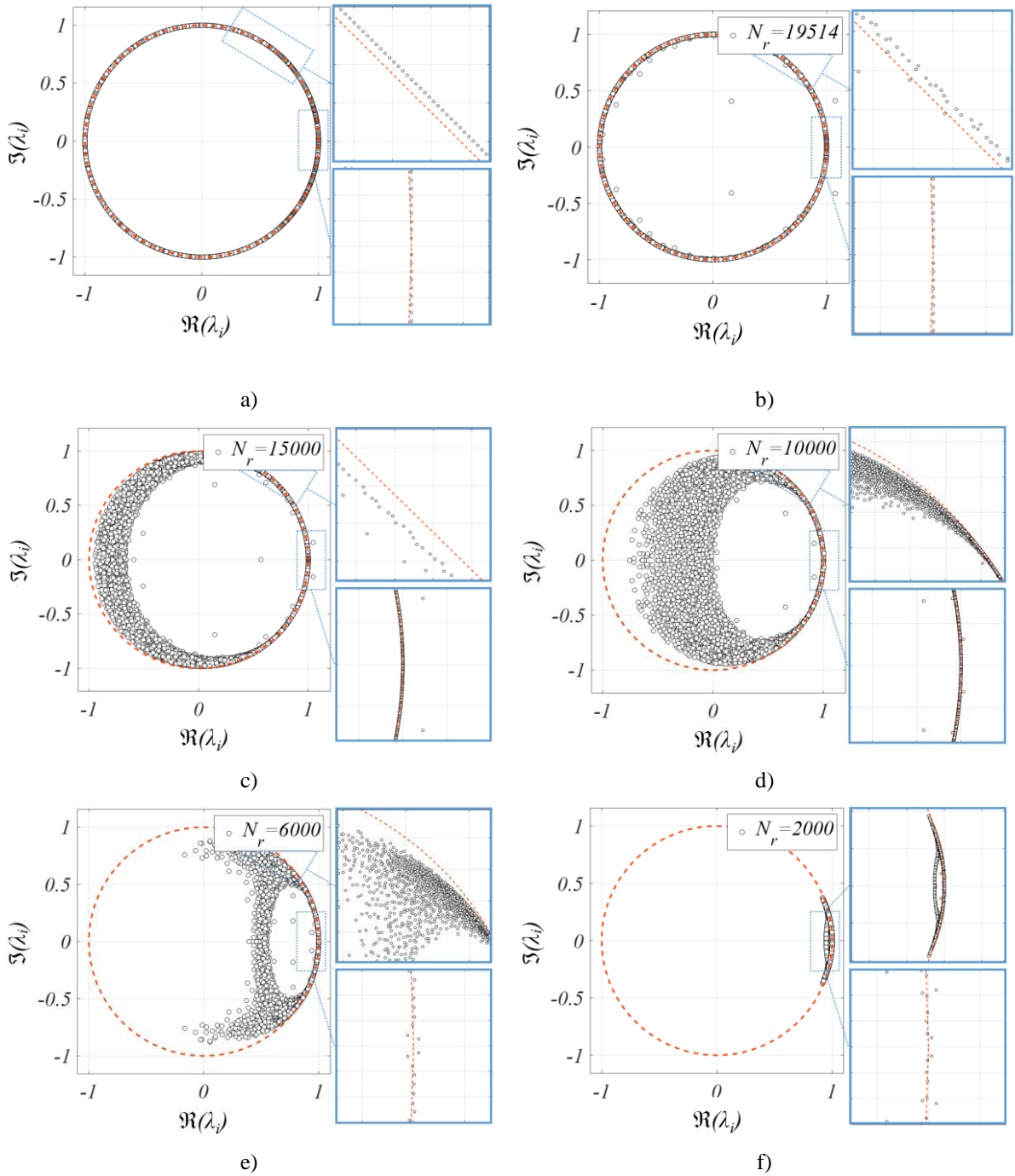

**Fig. 18** The DMD spectra of truncation order a) $N_r = 19515$ (full-order), b) $N_r = 19514$ c) $N_r = 15000$ d) $N_r = 10000$, e) $N_r = 6000$ f) $N_r = 2000$.

Finally, to settle the answered questions, we resort to the most indicative index of accuracy---reconstruction. In this case, we must impose a sufficiently large error cap of *100* to depict and distinguish divergence tangibly. **Fig.** 19 present the grand mean $l_2$-norm of reconstruction error, generalising the effect of truncation on reconstruction.

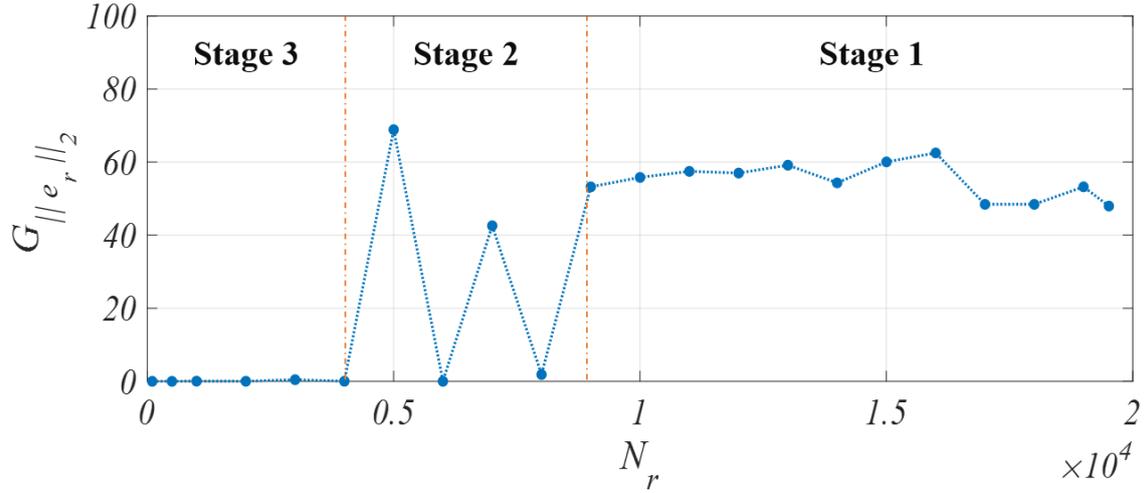

**Fig. 19** Grand mean $l_2$-norm of reconstruction error versus truncation order $N_r$ with maximum error threshold of *100*.

In combination with the observations from Section 5.2.2, the grand mean error lucidly depicts three distinct stages of truncation:

> **Stage 1** marks the range of large $N_r$ in which reconstruction of the original data is utterly flawed because of divergence. Incidentally, Stage 1 coincides with the range of $N_r$ in which the DMD preserves the dynamical integrity of the low-order modes.
>
> **Stage 2** marks the range of intermediate $N_r$ in which reconstruction experiences a transition. In this range, the error is small for some $N_r$ and diverges for others. The genesis of Stage 2 also coincides the loss of dynamical integrity of the lowest-order modes, indicating an utter shift of the spectral discretisation.
>
> **Stage 3** marks the range of small $N_r$ in which reconstruction becomes highly accurate and comparable to that of the full-order benchmark. However, the truncated systems yield modes of completely different dynamical characteristics.

Now, we can translate the empirical observations into a set of conclusions about truncation. Even by a single order, truncation weakens the stability and shifts the causality of the DMD output while causing divergence in data reconstruction. However, its effect on the low-order modes, or the most dominant descriptors of system dynamics, is limited with large $N_r$. As $N_r$ further decreases, the POD subspace and continuously shrink and the stability deteriorates, but the reconstruction accuracy restores after a demarcation. The demarcation implies the complete desertion of the original descriptors and is marked by the loss of temporal integrity of the lowest-order modes.

Finally, we briefly comment on the practicality of truncation as an order-reduction tool of the DMD and the Koopman analysis. If the preserving the Koopman eigen tuples is preferred, users shall refrain from truncation at all or, if unavoidable, to the best of their computational power. We also remind the readers that the stellar performance of the d.m.s., as revealed in [62], is established on a full-order POD subspace. Therefore, the peri-decomposition truncation will affect the performance of the post-decomposition d.m.s., too. On the other hand, if the investigative agenda is exclusively data reconstruction, we recommend a large-degree truncation for parsimonious computations.

## 5.3 Pre-Decomposition Interpolation

### 5.3.1 Pre- versus Post-Decomposition

We will dedicate the last section of the work to studying the effect of pre-decomposition interpolation. Our motivation arises from the need to avoid the *Divergence* state. One may interpolate the input data to increase the spatial dimension of *n* artificially, hence delaying the violation of the $m<n$ condition. Before we begin, readers may sympathise that interpolation is almost inevitable in the post-analysis fluid system, particularly in visualisations by contours or three-dimensional iso-surfaces. Unlike coding languages such as MATLAB or Python, interpolation is often automatically embedded in commercial software like ANSYS or Tecplot.

For this reason, we must distinguish the difference between pre- and post-decomposition interpolation. The deliberate or inadvertent interpolation for the purpose of visualisation is usually performed on the output post-analysis, therefore only interfering minimally with the algorithmic execution of whichever technique one deploys. It is beyond our intention to assess the adequacy of pre-decomposition interpolation or evaluate the accuracy of different schemes. We intend to answer the question: will, and if so, by how much will pre-decomposition interpolation affect the algorithmic execution of the DMD, or consequently, its output? Equivalently, one may think of the following discussion as a rudimentary parametric test for the feasibility of interpolation as a data pre-processing method of the DMD.

### 5.3.2 Stability, Information Retention, and Feasibility

We begin by assessing the stability of interpolated systems. The DMD spectra of the linearly and cubically interpolated sequences display different degrees of stability deterioration (**Fig.** 20). The cubic case evidently outperforms the linear case by preserving a greater portion of the original subspace, but both systems are causal and sampling dependent. As measured by the absolute value of the growth/decay rate, the deterioration quantifies to at least four to five orders of magnitude for both schemes (**Table** 5). Recalling Section 4.2, we conclude interpolation does not alter the spectral discretisation because $m$ is intact, but it adds new spatiotemporal information into the original system in the form of high-frequency noise with increases in $n$.

Scrutiny of the leading Koopman eigen tuples also reveals that the noise from the linear case overwhelms the original dynamical information, so the DMD output essentially becomes degenerate (**Table** 5). We failed to identify the original $M_{1-3}$ in the first 20 dominant modes for the linear case. Some noise of stationary features, like $St_1=0$, is also generated in a blatant contradiction to the original dynamics. On the other hand, though weakened in dominance (in green), the leading mode remains discernible and genuine in the allegedly more accurate cubic case. **Fig.** 21 verifies the better retention of information by illustrating the akin mode shapes of the original and cubically interpolated cases. To this end, we attribute the superior performance in stability and information retention to the cubic scheme's more truthful portrayal of the inter-nodal, nonlinear fluid dynamics. Though still far from ideal, the high-order scheme at least ensures that the original information is not overly contaminated, if not utterly smudged like the linear case.

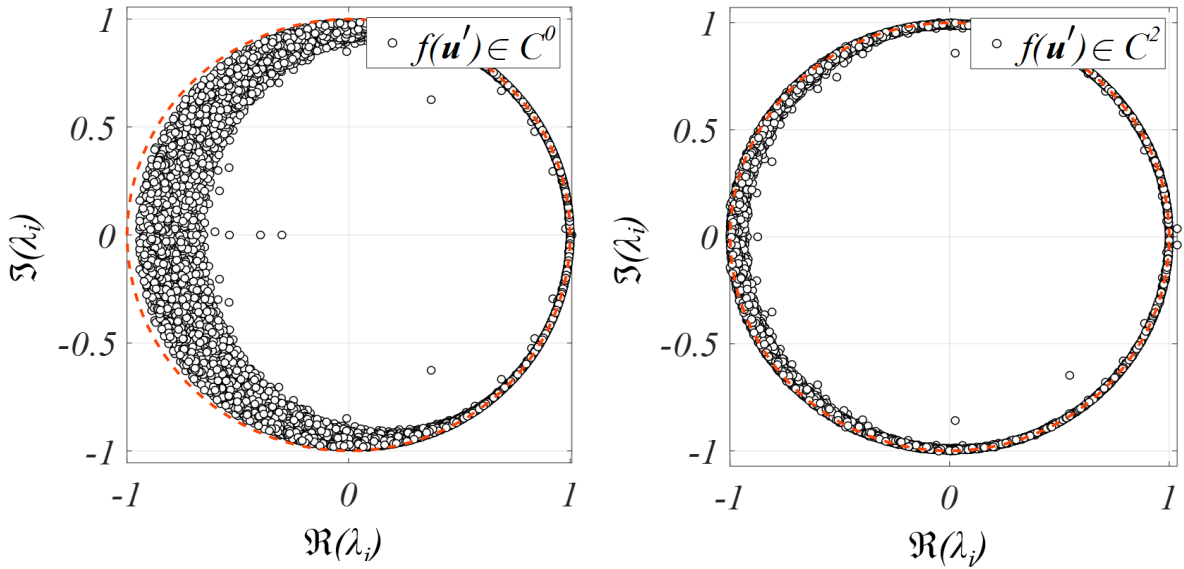

a)                        b)

**Fig. 20** The DMD spectra by sampling 20 oscillation cycles of fluctuating velocity with a) linearly interpolated data $f(\boldsymbol{u}')\in C^0$ and b) cubically interpolated data $f(\boldsymbol{u}')\in C^2$.

**Table 5** The Strouhal number $St$ and growth rate $g$ of DMD modes $M_{1-3}$ by sampling the original fluctuating velocity $\boldsymbol{u}'$, triangulation-based linearly interpolated $\boldsymbol{u}'$, $f(\boldsymbol{u}')\in C^0$, and triangulation-based cubically interpolated $\boldsymbol{u}'$, $f(\boldsymbol{u}')\in C^2$.

|    |       | $\boldsymbol{u}'$ | $f(\boldsymbol{u}')\in C^0$ | $f(\boldsymbol{u}')\in C^2$ |
|----|-------|---------|---------|---------|
|    | $M_1$ | 0.128   | 0       | 0.727   |
| $St$ | $M_2$ | 0.121   | 0.594   | 1.46    |
|    | $M_3$ | 0.00638 | 0.737   | 0.128   |
|    | $M_1$ | $-1.99\times 10^{-7}$ | $7.72\times 10^2$ | $3.63\times 10^3$ |
| $g$ | $M_2$ | $-1.77\times 10^{-7}$ | $-2.05\times 10^3$ | $-9.36\times 10^2$ |
|    | $M_3$ | $-1.34\times 10^{-7}$ | $-2.03\times 10^2$ | $1.97\times 10^{-2}$ |

At this point, we must emphasise that any pre-decomposition interpolation, regardless of precision or integrity, adds synthetic information into the original data, which will manifest as noise after spectral characterisation. The synthetic noise is inevitable unless the interpolation scheme is precisely the Navier-Stokes equations, which is unrealistic. The level of noise is also hard to quantify even for reconstruction because the original data has been modified.

In practice, if pre-decomposition interpolation is unavoidable, say when field or wind tunnel measurement only captures very limited data due to apparatus limitation, we recommend users to deploy high-order schemes for better retention of the original dynamics. Nevertheless, we remind the readers about our vain attempt to apply the purportedly more accurate bi-harmonic spline to the current data. The ambitious trial led to the overheating of our servers' RAM and CPU for weeks, before we eventually capitulated afront the prohibitive computational cost. Finding an ideal balance between cost and quality has always been and will forever be one of the greatest dilemmas of research, if not life in general.

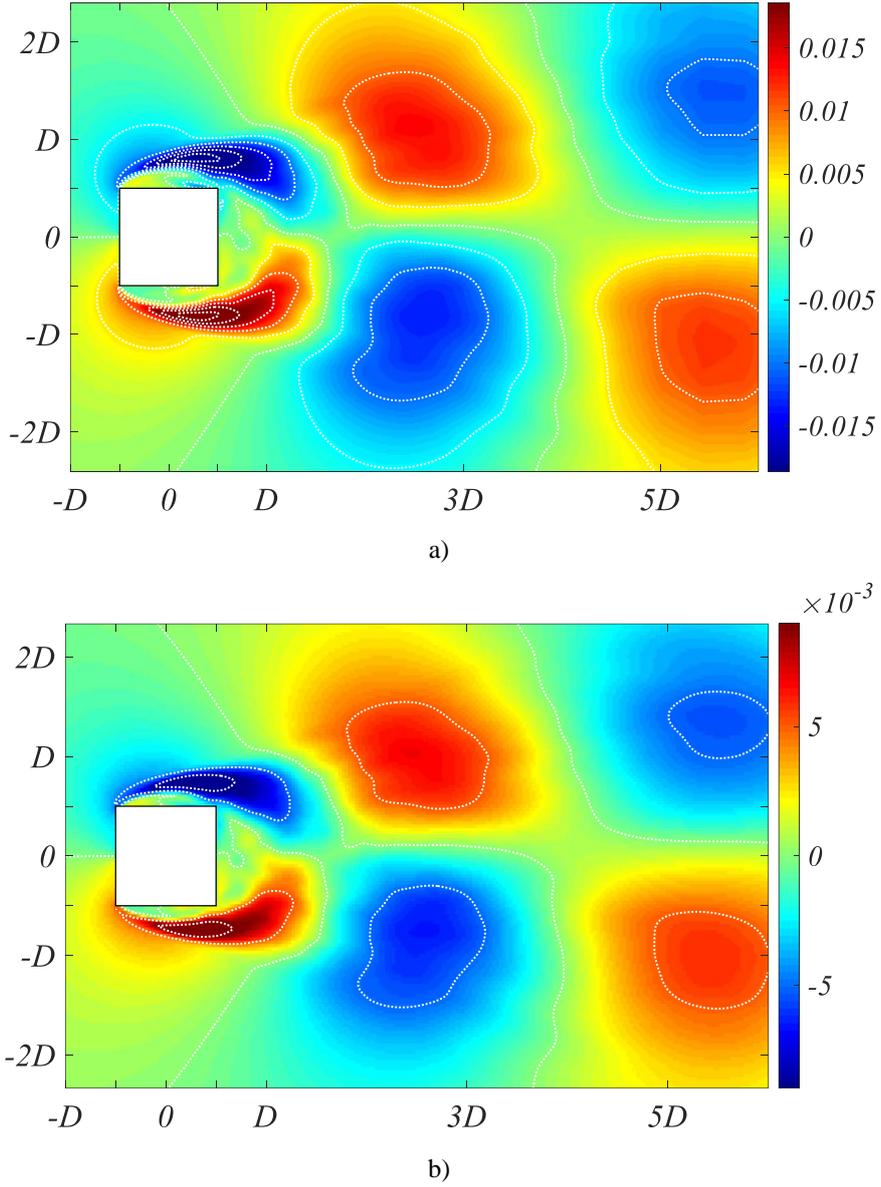

**Fig. 21** Mode shape $\Re(\phi_3)$ of oscillatory DMD modes $M_3$ by sampling 20 oscillation cycles of a) fluctuating velocity $\boldsymbol{u}'$ and b) $\boldsymbol{u}'$ interpolated on an equidistant grid ten times larger by the cubic triangulation-based convolution scheme $f(\boldsymbol{u}') \in C^2$.

## 6. Conclusions

On a conclusive note, we summarise the merit of this parametric work on the sampling nuances of the DMD and the Koopman analysis. Through the lens of turbulent wakes, the work first corroborated the generality of the universal convergence states of DMD sampling. Subsequently, it discovered the spectral implications of the sampling range, or the temporal dimension $m$, which controls the resolution of the spectral discretisation and globally dictates the width of the equidistant frequency bins. On the other hand, the sampling resolution, or the inter-snapshot step size, controls the upper limit of the spectrum, so the highest resolved frequency. The spectral implications suggest that the convergence states are *de facto* universal for DMD implementation, so the arrival of the *Convergence* state is essential for sampling independence.

Afterwards, the present work parametrically examined the effect of input observables. Results from 18 observables suggested highly dependent DMD output because the variables, though arising from the same flow, contain vastly different dynamical information. Therefore, selecting appropriate observables is of prime importance for successful DMD efforts. Wall pressure and eigenvalue-based or ratio-based vortex identification criteria are excellent indices for structural response and fluid excitation. The pressure, velocity magnitude, and turbulence kinetic energy fields also suffice for general applications, but the Reynolds stresses and velocity components shall be avoided without in-depth knowledge about a fluid system. Furthermore, mean-subtraction is recommended as the inclusion of the mean-field substantially weakens the stability and temporal integrity of the Koopman approximations.

The present work also parametrically assessed the effect of truncation, discovering the low-energy states that are trivial to a system's POD energy but dictate its temporal integrity. Results also showed that truncation, even by a single order, weakens the stability and shifts the causality of the output, while causing divergence in data reconstruction. In practice, truncation shall be avoided for high-fidelity approximations of the Koopman eigen tuples. The best practice for order reduction, given sufficient computational resources, is to employ dominant mode selection on a full-state subspace. Nevertheless, if data reconstruction is the sole purpose, large-degree truncation fulfils the job at a low computational cost.

This parametric work also answers the outstanding question about interpolation. Pre-decomposition interpolation inevitably introduces synthetic noise into a system's intrinsic dynamics, hampering the decomposition integrity. If interpolation is compulsory, for example, to increase the spatial dimension $n$ to meet the $m<n$ condition, we recommend using high-order schemes for better retention of the original dynamics. In sum, the parametric work fills the gap in knowledge by offering several application-oriented guidelines for the DMD and the Koopman analysis. Observations from the inhomogeneous anisotropic turbulent wakes may also extend to a wide array of fluid systems, perhaps even some nonlinear systems beyond fluid mechanics.

# Acknowledgements

We give a special thanks to the IT Office of the Department of Civil and Environmental Engineering at the Hong Kong University of Science and Technology. Its support for installing, testing, and maintaining our high-performance servers is indispensable for the current project.

## Funding

The work described in this paper was supported by the Research Grants Council of the Hong Kong Special Administrative Region, China (Project No. 16207719), the Fundamental Research Funds for the Central Universities of China (Project No. 2021CDJQY-001), the National Natural Science Foundation of China (Project No. 51908090 and 42175180), the Natural Science Foundation of Chongqing, China (Project No. cstc2019jcyj-msxmX0565 and cstc2020jcyj-msxmX0921), the Key project of Technological Innovation and Application Development in Chongqing (Project No. cstc2019jscx-gksbX0017), and the Innovation Group Project of Southern Marine Science and Engineering Guangdong Laboratory (Project No. 311020001).

## Conflict of Interest

The authors declare that they have no conflict of interest.

## Availability of Data and Material

The datasets generated during and/or analysed during the current work are restricted by provisions of the funding source but are available from the corresponding author on reasonable request.

## Code Availability

The custom code used during and/or analysed during the current work are restricted by provisions of the funding source.

## Author Contributions

All authors contributed to the study conception and design. Funding, project management, and supervision were led by Tim K.T. Tse and Zengshun Chen and assisted by Xuelin Zhang.

Material preparation, data collection, and formal analysis were led by Cruz Y. Li and Zengshun Chen and assisted by Asiri Umenga Weerasuriya, Yunfei Fu, and Xisheng Lin. The first draft of the manuscript was written by Cruz Y. Li and all authors commented on previous versions of the manuscript. All authors read, contributed, and approved the final manuscript.

## Compliance with Ethical Standards

All procedures performed in this work were in accordance with the ethical standards of the institutional and/or national research committee and with the 1964 Helsinki declaration and its later amendments or comparable ethical standards.

## Consent to Participate

Informed consent was obtained from all individual participants included in the study.

## Consent for Publication

Publication consent was obtained from all individual participants included in the study.

# Appendix I

The truncated materials in this section are taken from our previous work [16] for the convenience of readers.

*Statistical Stationarity*

**Global Statistics**

The fluctuating lift coefficient is presented in **Fig.** AI.1 for global stationarity. The root-square-mean (r.m.s.) and the mean lift reached statistical stationarity before $1.2 \times 10^5 t^*$, so sampling began thereafter. The sample contains *24* oscillation cycles at the highest frequency $1/t^*$, pushing the storage limit of our HPC server by amassing more than $2.3 \times 10^4$ field snapshots.

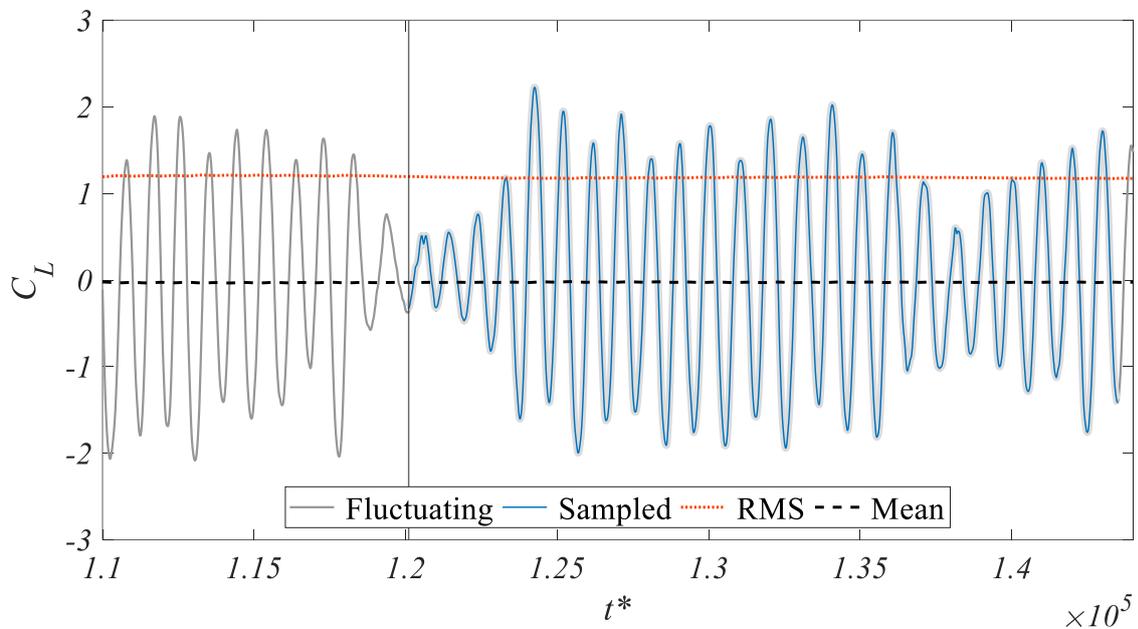

**Fig. AI.1** Time histories of instantaneous, RMS, and mean lift coefficients. The DMD sampling range consists of 24 oscillation cycles in the statistical stationary state.

**Local Statistics**

The statistical stationarity is reinforced by local statistics. **Fig.** AI.2 illustrates the seven nodes selected to monitor the mean local velocities. At all monitor points, the normalized mean velocities exhibited stationarity by approaching clear asymptotes before $1.0 \times 10^5 \; t^*$, reaffirming the observations on global statistics.

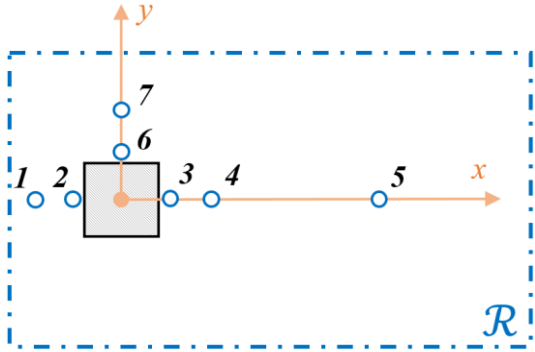

a)

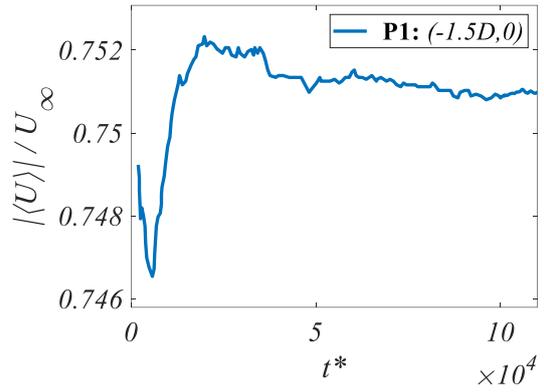

b)

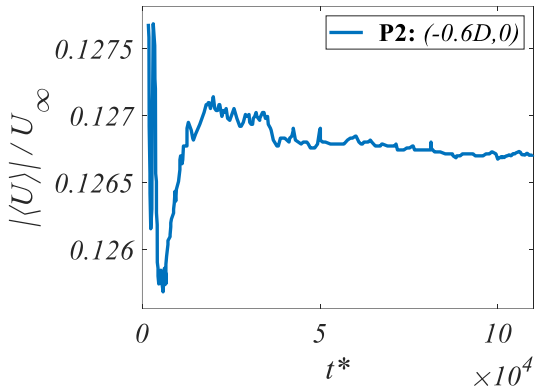

c)

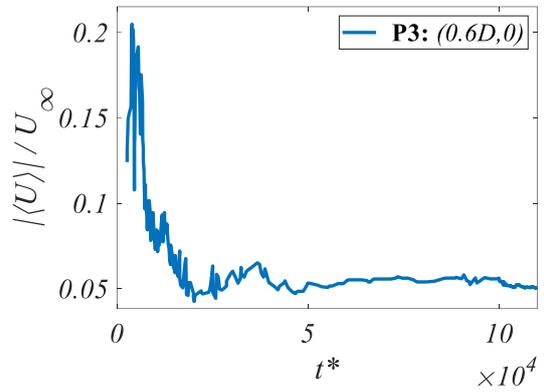

d)

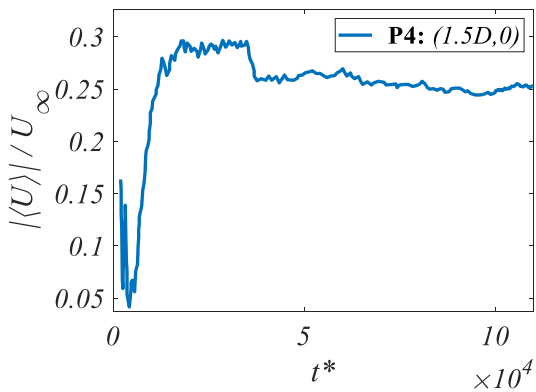

e)

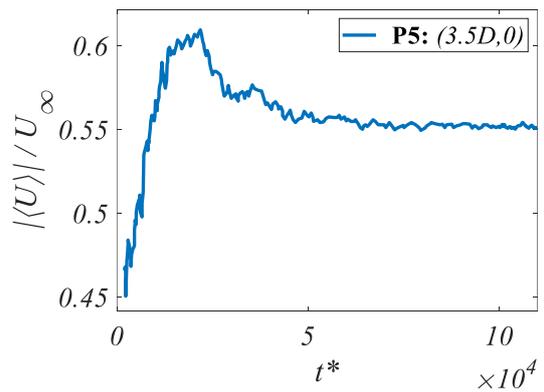

f)

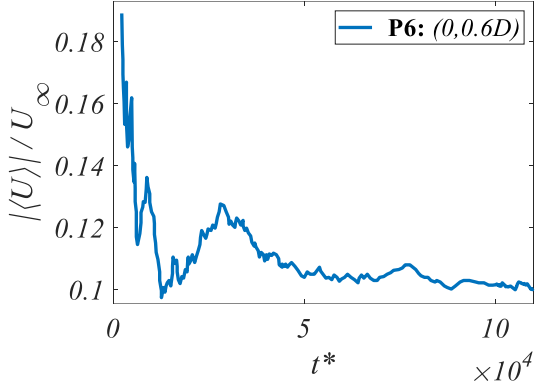
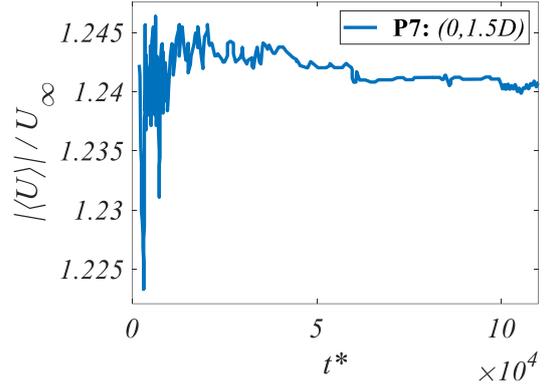

g)  h)

**Fig. AI.2** a) A schematic illustration showing the locations of monitor points 1-7; b)-h) normalized mean velocity magnitude versus time step at monitor points 1-7, showing statistical stationarity.

### *Grid Resolution*

The resolution of the structured grid $G$ is critically important to the accuracy of LES-NWR, so stringent criteria were adopted for its assessment.

**Resolved Spectrum**

A proper LES-NWR shall have its filter sitting inside the inertial subrange, so the Navier-Stokes equations guarantee the accuracy of large-scale turbulence. Pope [30] defined the demarcation between the energy-containing range and the inertial subrange -- 80% resolution for the total turbulence kinetic energy $\langle k \rangle$.

To quantify the resolved portion, the resolved portion of $\langle k \rangle$ is denoted by $k_r$, the subgrid portion by $k_{sgs}$, and the numerical portion by $k_{num}$. The resolved spectrum, $E$, is expressed by

$$E \equiv G(x)\, E_{\langle k \rangle} \approx \frac{k_r}{\langle k \rangle}, \qquad (AI.1)$$

where

$$\langle k \rangle = k_r + k_{sgs} + k_{num}, \qquad (AI.2)$$

$$k_r = \frac{1}{2}\left(\overline{u'^2} + \overline{v'^2} + \overline{w'^2}\right), \qquad (AI.3)$$

$$k_{sgs}=v_{sgs}^2/l_s^2, \qquad (AI.4)$$

where $G(x)$ denotes the filter function in three-dimensional space; $\overline{u'^2}, \overline{v'^2}$, and $\overline{w'^2}$ denote variance of the fluctuating velocities $u'$, $v'$, and $w'$, respectively. $k_{num}$ is a pseudo-energy term that accounts for discretisation error and numerical residual. Celik *et al.* [64] pointed out that $k_{num}$ is sufficiently small for a LES-NWR with an overall second-order discretisation. As introduced before, our discretisation is at least second-order, producing minimal numerical dissipation. As the Courant-Friedrichs-Lewy (*CFL*) condition is strictly maintained, the numerical dispersion is insignificant. Our convergence criteria are also stringent, at $1\times10^{-6}$, for both the continuity and momentum equations. Therefore, $k_{num}$ is treated as negligible.

The ensuing **Fig.** AI.3 presents the resolved spectra of the mid-span x-y plane and the prism walls. Evidently, the grid resolved at least 90% of the $\langle k \rangle$. The same was true for the prism walls, except for some narrow strips near the corners A and B, where the resolved $\langle k \rangle$ is about 77%. They are anticipated because local accelerations occur as the result of sharp corner separation. The resolved spectra prove that our filter sits inside the inertial subrange, hence the resolution of our grid is suitable for LES simulation.

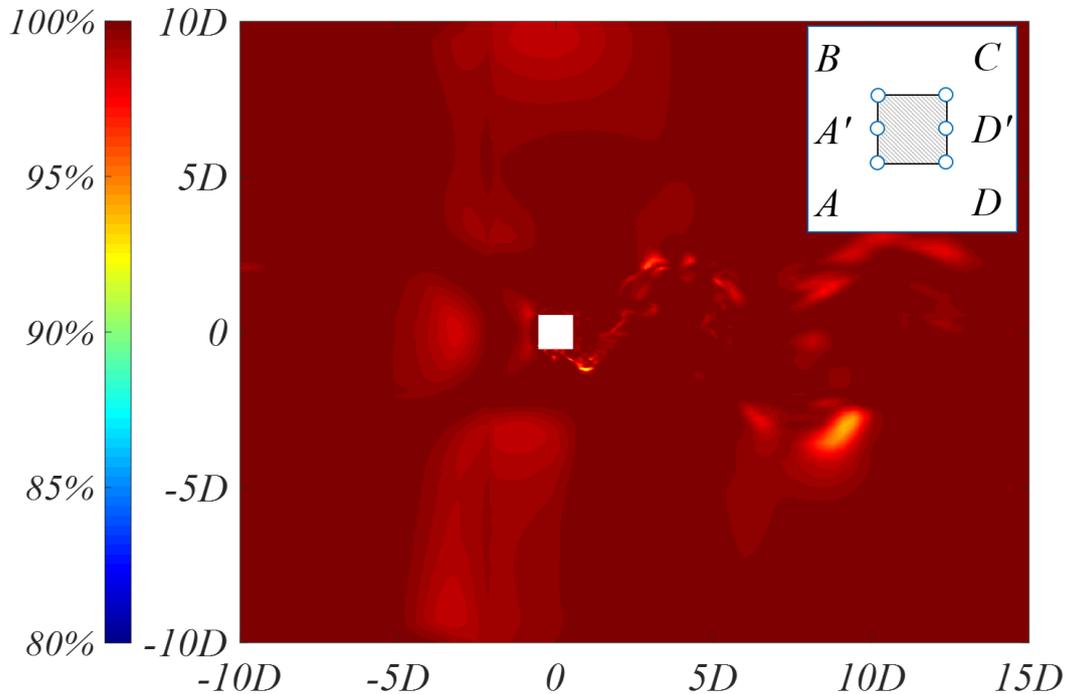

a)

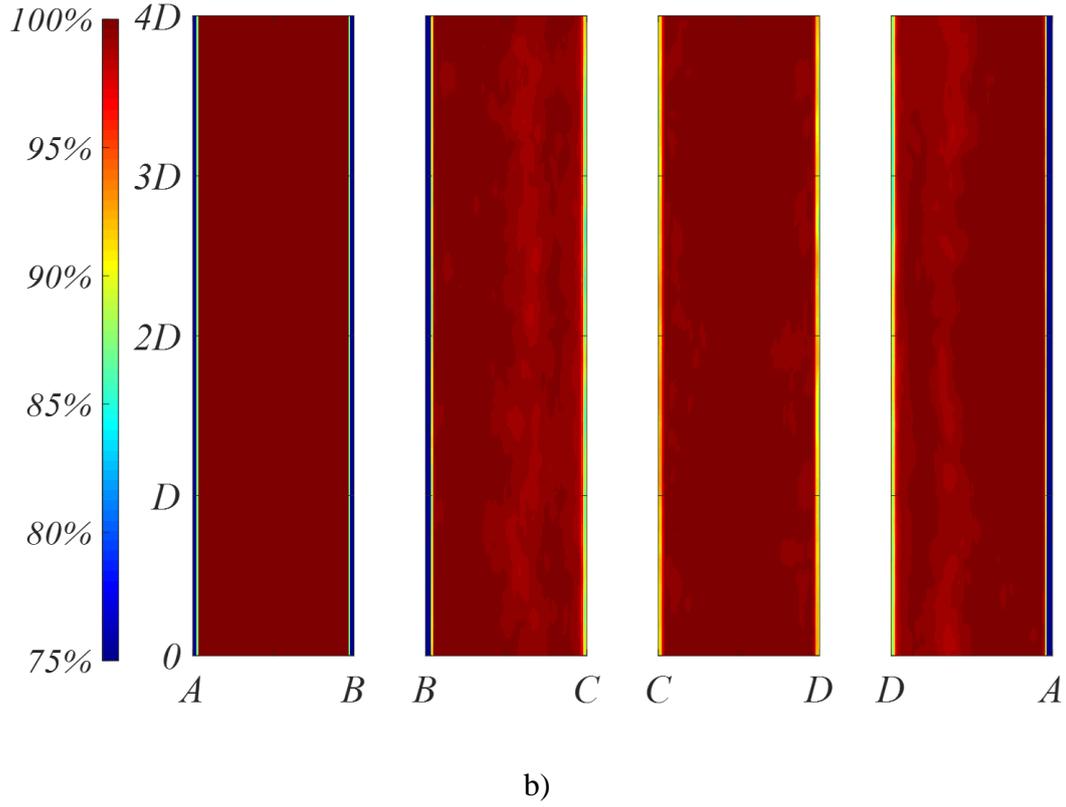

**Fig. AI.3** Contours showing the time-averaged resolved spectra of the turbulence kinetic energy of a) the mid-span x-y plane and b) prism walls. 80% resolves at least the energy-containing range.

**LES$_{IQ}$ Index**

In the fluid mechanics community, some researchers, like Davidson [65], had challenged the 80% demarcation. Though this topic is beyond the current scope, an alternative assurance is still provided to corroborate the grid resolution. Celik *et al.* [64] proposed the *LES$_{IQ}$*, an index quantifying the resolution of a LES grid, as

$$LES_{IQ} = \left[1 + a_v \left(\frac{v_{sgs} + v}{v}\right)\right]^{-a_e}, \qquad (AI.5)$$

where $\alpha_v = 0.05$ and $a_e = 0.53$ are empirically derived constants. Accordingly, *LES$_{IQ}$ < 80%* indicates a Very Large-Eddy Simulation (VLES), *LES$_{IQ}$ > 80%* signals a proper LES-NWR, and *LES$_{IQ}$ > 95%* suggests a Direct Numerical Simulation (DNS).

The ensuing **Fig.** AI.4 presents the *LES$_{IQ}$* contours of the mid-span x-y plane and the prism walls. The grid achieved the DNS resolution in most fluid domains and on all prism walls. It also achieved at least the LES resolution in all fluid domains of interest, including the shear layers and the near wake. Although a few patches of VLES resolution existed in the far wake but being neither influential to the upstream activities nor partial to the sampled domain, they are of trivial importance. Overall, the $LES_{IQ}$ confirms the quality of the grid resolution.

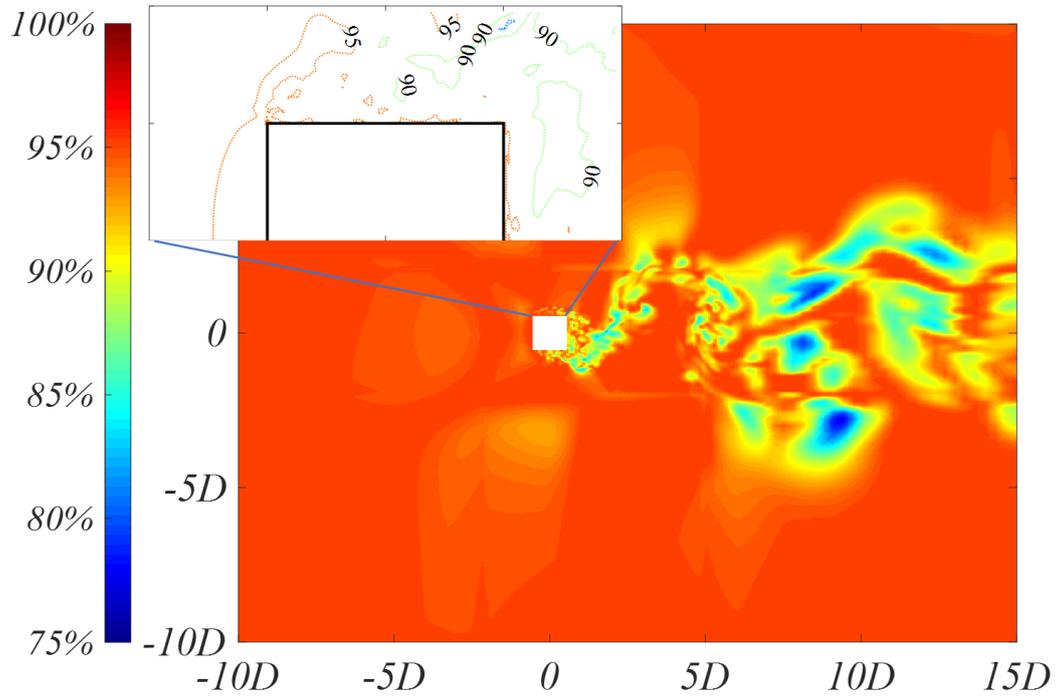

a)

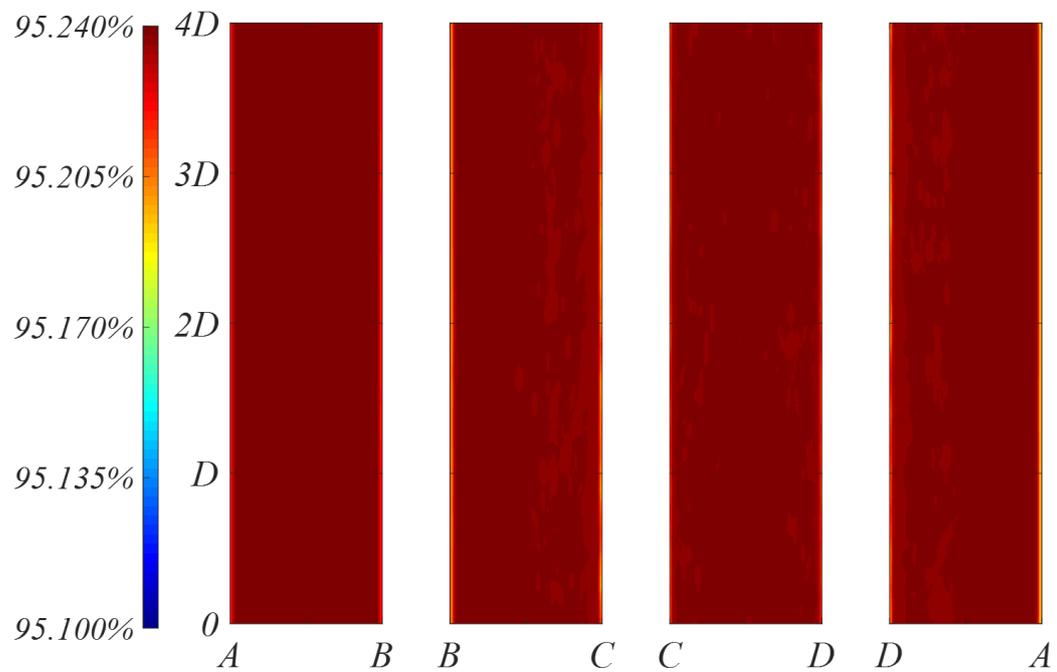

b)

**Fig. AI.4** Contours showing the $LES_{IQ}$ index proposed by Celik *et al.* [64] of a) the mid-span x-y plane and b) prism walls. 80% signals satisfactory LES resolution.

*Simulation Accuracy*

Since numerical accuracy is indispensable to the subsequent DMD analysis, beside the grid resolution, we compare the simulated results to the literature for further validation.

**Global Statistics and Spectral Density**

Below, **Table** AI.1 summarizes global force coefficients and the Strouhal number *St* of the present simulation compared to previous experiments. The statistics not only agree with the experiments performed at exactly *Re=22,000,* but also an array of others in the subcritical regime, because one expects the universal occurrence of the shear layer transition II and an asymptotic convergence vortex formation length in this flow regime [18].

**Table AI.1** Global force coefficients and the Strouhal number compared to the literature

| $Re$ $(10^3)$ | $\langle C_D \rangle$ | $C_{D, RMS}$ | $C_{L, RMS}$ | $St$ | Contribution |
|---|---|---|---|---|---|
| **22** | **2.048** | **0.200** | **1.173** | **0.127** | **Present work** |
| 22 | 2.069 | 0.146 | 1.221 | 0.126 | Li *et al.* [62] |
| 100 | 2.05 | 0.17 | 1.3 | 0.12 | Vickery [66] |
| 176 | 2.04 | 0.22 | 1.19 | 0.122 | Lee [67] |
| 22 | 2.1 | - | 1.2 | 0.13 | Bearman/Obasaju [68] |
| 27 | 1.9-2.1 | 0.1-0.2 | 0.1-0.6 | - | Cheng *et al.* [69] |
| 23 | 1.9-2.1 | 0.1-0.2 | 0.7-1.4 | - | McLean/Gartshore [70] |
| 22 | 2.10 | - | - | 0.130 | Norberg [71] |
| 34 | 2.21 | 0.18 | 1.21 | 0.13 | Luo *et al.* [72] |
| 21.4 | 2.1 | - | - | 0.132 | Lyn *et al.* [73], [74] |

On the other hand, the Strouhal number *St* is determined by the power spectral density analysis for the fluctuating global lift coefficient, $C_L{}'$. Apart from the stellar agreement in global *St*, the periodogram (**Fig.** AI.5) supports previous conclusions of the grid resolution. The curve lucidly exhibits the Kolmogorov *-5/3* Law before *St=1.1*. As a fundamental pillar to the Richardson-Kolmogorov energy cascade, the -5/3 power-law is unique to the inertial subrange and often regarded as its most indicative icon [30]. Therefore, its appearance signifies that at least a significant portion of the inertial subrange was resolved by the Navier-Stokes equations. On a different note, one shall expect an exponential decay in the dissipation range of the full turbulence spectrum, instead of the linear decay herein towards the high-frequency space. The observed linearity is truthful to the subgrid dynamics, especially considering the one-equation, linear mixing-length hypothesis undertaken by the Smagorinsky model. The linearity shows the filtering process only takes place inside and beyond the inertial subrange, depicting the anticipated spectrum of a proper LES-NWR.

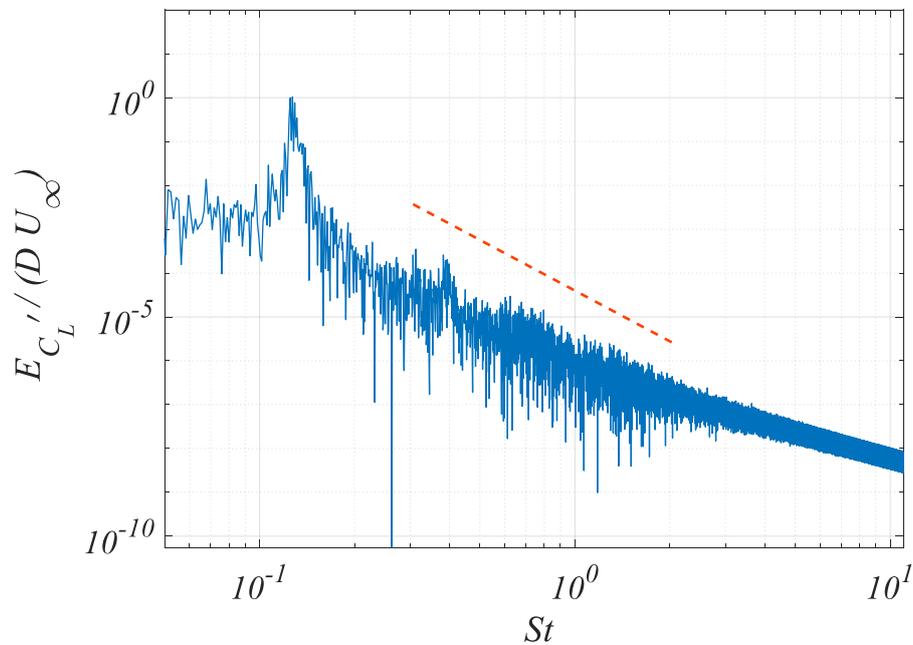

**Fig. AI.5** Periodogram of the normalized spectrum of the fluctuating global lift coefficient, showing the Strouhal number for the Kármán vortex shedding and an agreement with the Kolmogorov -5/3 Law [30].

**Prism Walls**

Next, the prism walls are examined for validation. **Fig.** AI.6 presents the $x^+$ and $y^+$, or the dimensionless wall distances, calculated from wall shear and the notion of friction velocity. The grid met the $x^+\approx30$ and $y^+\approx1$ requirements proposed by Menter [75]. The $z^+$ is trivial in this infinite length configuration. Moreover, the time-averaged, normalized pressure coefficient on the prism walls is compared to the literature (**Fig.** AI.7). The present results agree well with a range of wind tunnel and DNS studies, especially the more recent empirical results like Nishimura [76] and Nishimura and Taniike [77]. Notably, some variations occur on the downstream wall across the existing studies. Nonetheless, the present result falls well within the comparative range of wind tunnel and DNS studies, validating the accuracy and fidelity of the LES-NWR.

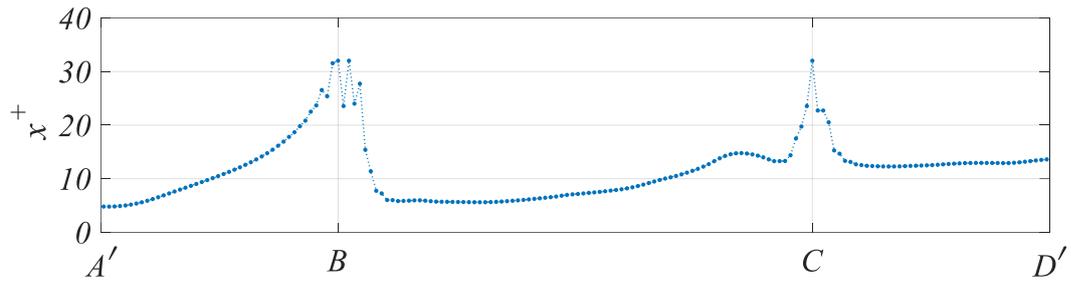

a)

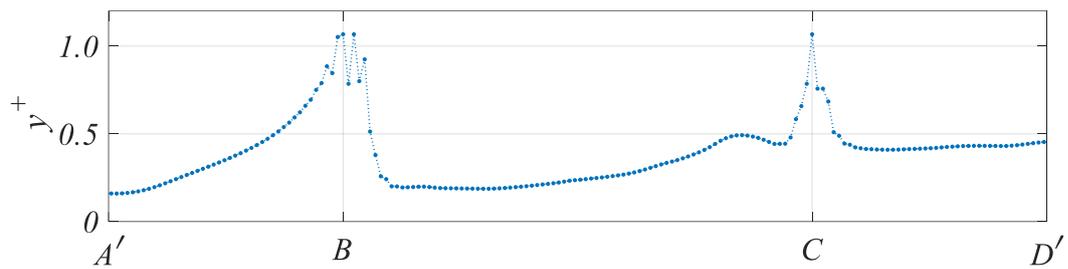

b)

**Fig. AI.6** Time-averaged values of a) $x^+$ and b) $y^+$.

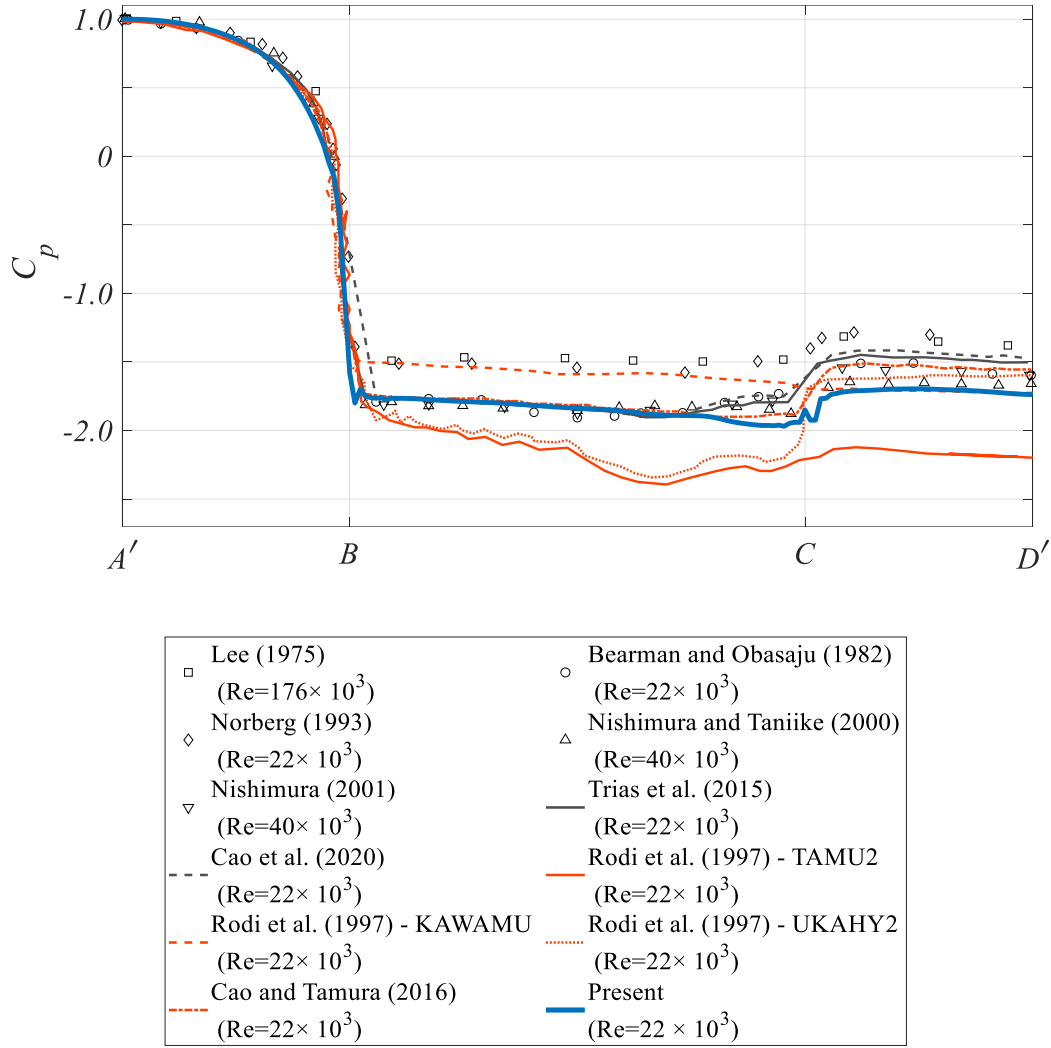

**Fig. AI.7** Time-averaged and normalized pressure coefficient of prism walls compared to the literature. Markers: experiments. Curves: black-DNS, orange-LES, blue-present work.

**Velocity Field**

The velocity field is also examined. **Fig.** AI.8 presents the time-averaged, normalized *u* along the zero-ordinate. The results sit fittingly among the array of experiments and DNS studies. To ensure such an agreement is not fortuitous, the fluctuating velocities that characterise turbulence are also verified. **Fig.** AI.9 presents the time-averaged, normalized *u', v',* and *w'* along the zero-ordinate. The agreement between the present results and the literature, especially the DNS results, is self-evident.

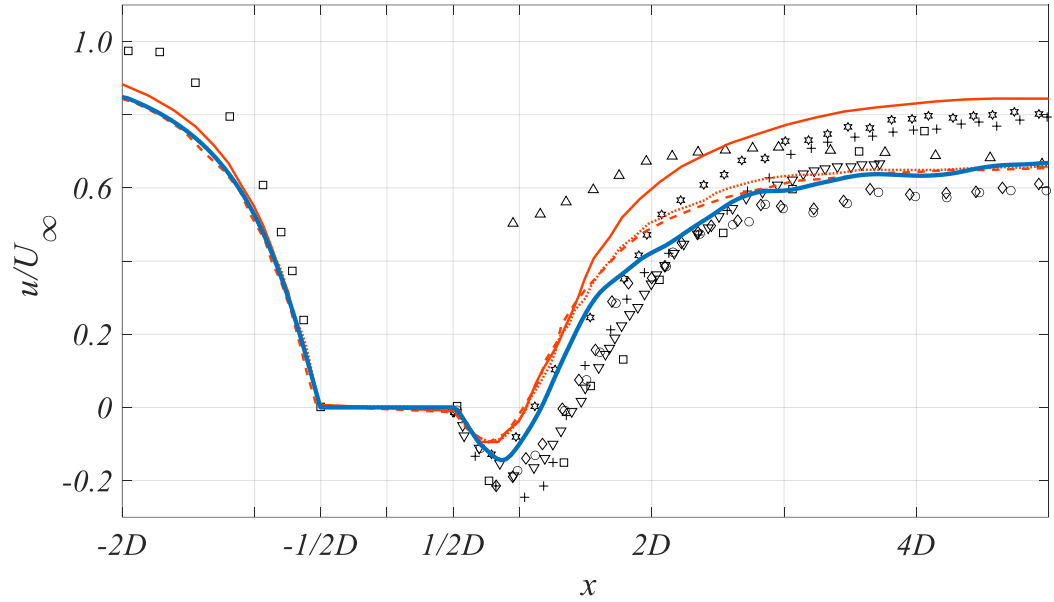

**Fig. AI.8** Time-averaged and normalized *u* along the zero-ordinate compared to the literature. Markers: experiments. Curves: orange-DNS, blue-present work

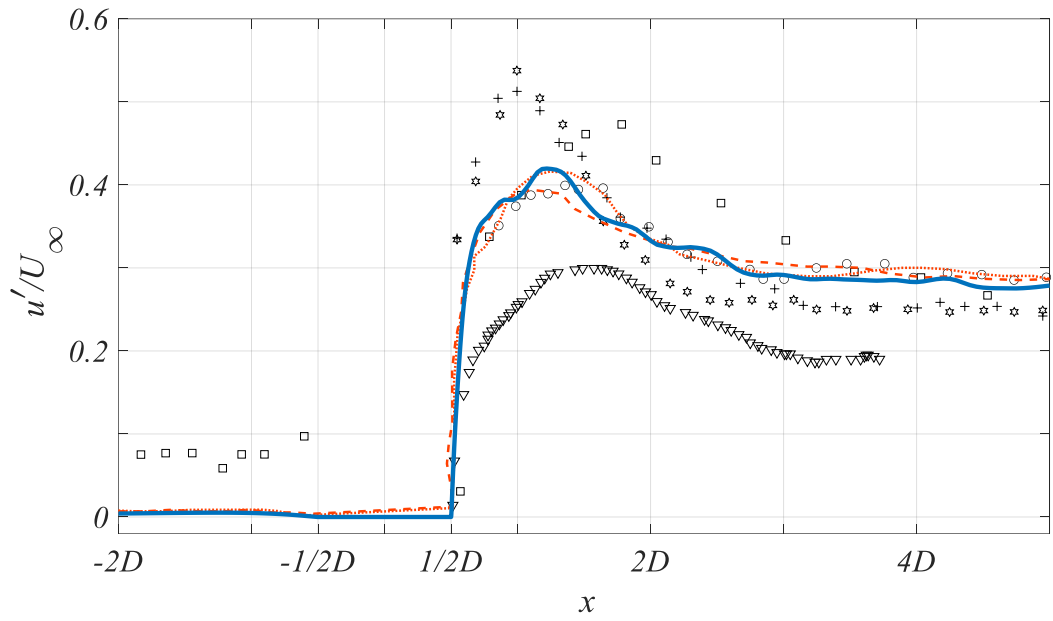

a)

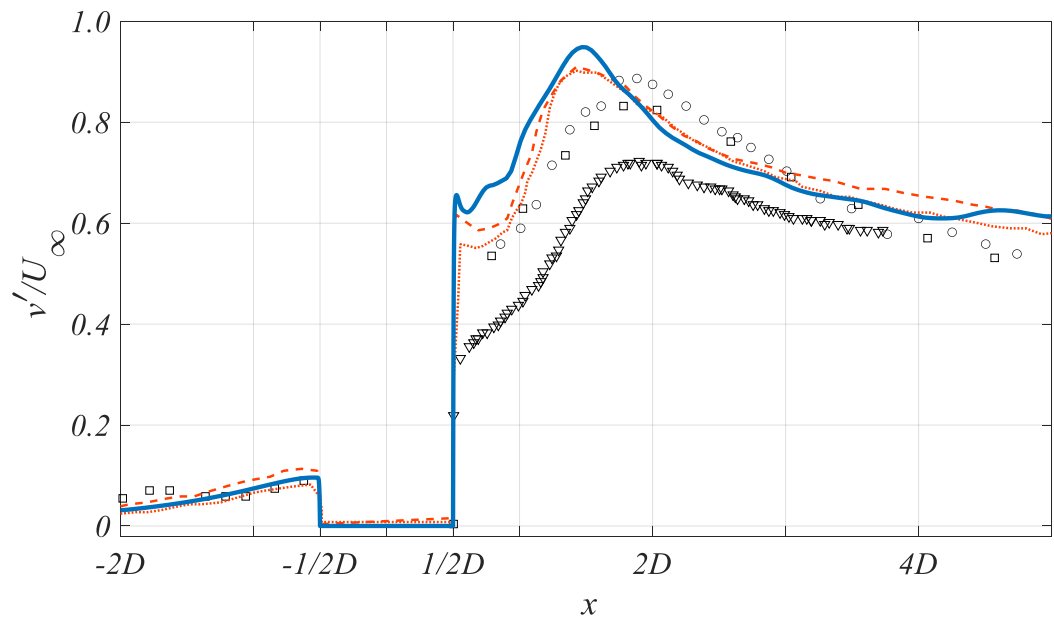

b)

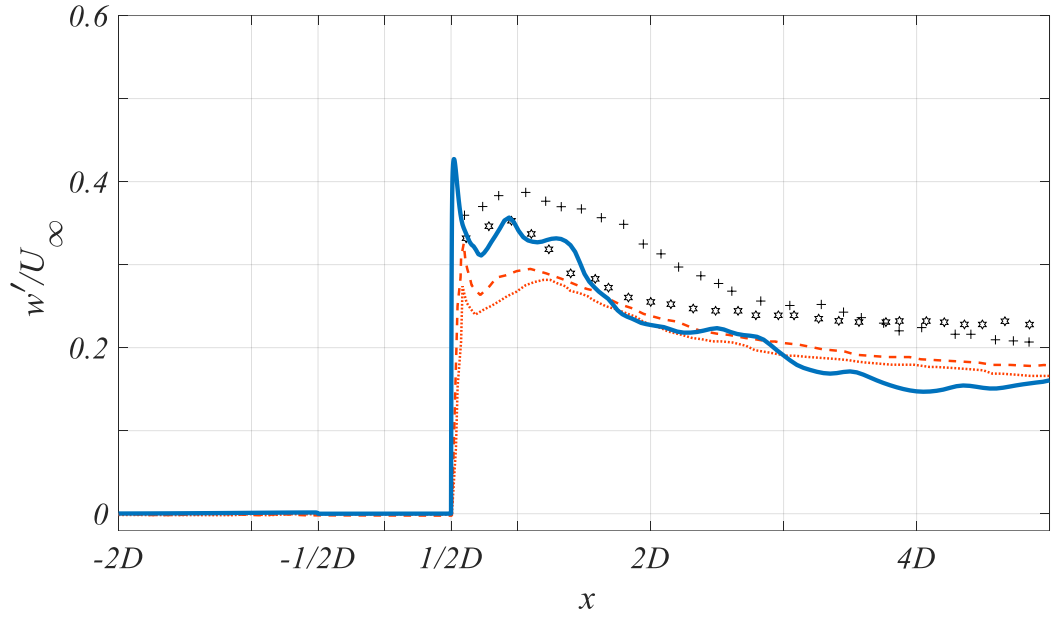

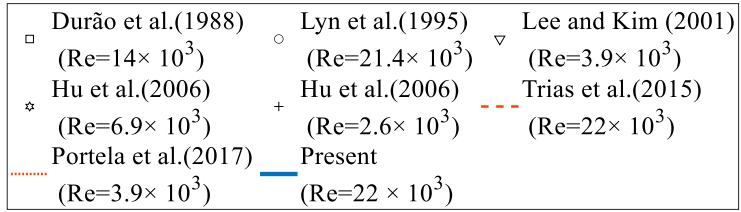

c)

**Fig. AI.9** Time-averaged and normalized fluctuating a) *u*, b) *v*, and c) *w* along the zero-ordinate compared to the literature.

# Appendix II

*Mathematical Formulation of the Similarity-Expression Dynamic Mode Decomposition*

Suppose one has a snapshot sequence of a particular variable of interest, or observable, sampled from a fluid system. The sequence can be arranged into two matrices, $X_1$ and $X_2$, that are separated by a uniform time step,

$$X_1 = \{x_1, x_2, x_3, \ldots, x_{m-1}\} \tag{AII.1}$$

$$X_2 = \{x_2, x_3, x_4, \ldots, x_m\}, \tag{AII.2}$$

where $x_i \in \mathbb{C}^n$ are individual data snapshots in the vector form. $n$ denotes the spatial dimension of the data sequence, which is capped by the maximum spatial resolution of the measurement apparatus or numerical grid. This work considers a fixed spatial domain. Readers may refer to Erichson *et al.* [78] for techniques dealing with an unfixed spatial domain. $m$ denotes the temporal dimension of the data sequence, which is capped by the sample size of the data sequence. Reader may refer to Kutz *et al.* [11] for techniques dealing with non-uniform sampling. Additionally, readers are reminded of a tacit assumption of the DMD, $m \ll n$.

The beauty of the DMD exudes precisely from the path it provides to forthrightly compute an approximation of the infinite-dimensional Koopman operator on a finite-dimensional subspace. The key is to assume a mapping matrix $A$ that connects $X_1$ and $X_2$,

$$X_2 = AX_1. \tag{AII.3}$$

Evidently, $A$ is an unknown matrix that mimics the map $f$, hence that of the Koopman operator $U$. Intuitively, the accuracy of $A$ increases with the dimensionality, so too is the computational expense to obtain $A$.

By the similarity-matrix expression, one may impose another matrix, the so-called similarity matrix $\tilde{A}$. The imposition of $\tilde{A}$ provides a computationally straightforward and algebraically tractable way to approximate $A$ by a singular-value-based algorithm. Readers may find the mathematical derivation and algorithmic procedure from Tu *et al.* [37] and Kutz *et al.* [11]. Sparing repetition, this section only outlines the plainest mathematics deployed in this implementation.

A Singular-Value Decomposition (SVD) is performed on $Y_1$,

$$X_1 = U\Sigma V^T, \qquad (AII.4)$$

where $U \in \mathbb{C}^{n \times r}$ contains spatially orthogonal Proper Orthogonal Decomposition (POD) modes $u_j$ on an optimal subspace; $\Sigma \in \mathbb{C}^{r \times r}$, a diagonal matrix, contains singular values $\sigma_j$ that describe the energy of $u_j$; $V \in \mathbb{C}^{m \times r}$ contains temporally orthogonal modes $v_j$ pertaining to the evolution of $u_j$; the superscript $^T$ denotes the conjugate transposition; $r$ denotes the truncation rank, which controls the order of $\tilde{A}$.

The POD-projected $\tilde{A} \in \mathbb{C}^{r \times r}$ relates to $A$ by

$$A = U\tilde{A}U^T. \qquad (AII.5)$$

Minimizing the difference between $X_2$ and $AX_1$, one obtains the expression

$$\underset{A}{\text{minimize}} \; \|X_2 - AX_1\|_F^2, \qquad (AII.6)$$

where $\| \; \|_F$ denotes the Frobenius normalization. Then, equations (A.4) and (A.5) are substituted into equation (A.6) in place of $X_1$ and $A$, respectively, obtaining

$$\underset{\tilde{A}}{\text{minimize}} \; \|X_2 - U\tilde{A}\Sigma V^T\|_F^2. \qquad (AII.7)$$

According to Tu *et al.* [37], the expression reduces to the approximation

$$A \approx \tilde{A} = U^T X_2 V \Sigma^{-1}. \qquad (AII.8)$$

Therefore, one may write the similarity-matrix-approximated expression of equation (A.3) as

$$X_2 = \tilde{A} X_1, \qquad (AII.9)$$

in which $\tilde{A}$ is the data-driven, reduced-order, and globally optimal approximation of $A$, hence the Koopman operator $U$. An attractive feature of this approach is that the deduction of $\tilde{A}$ is exclusively implicit, meaning it does not require explicit knowledge of a system's underlying dynamics, which is particularly useful for fluid analysis.

Intuitively, an equivalent data-driven approximation $\widetilde{\mathcal{A}}$ for the universal Koopman operator $\mathcal{U}$, as well as the finite-dimensional $\tilde{\mathcal{U}}$, also exists. The *a posteriori* observations deduced from the fluid-structure system herein, as will be presented in the subsequent sections, allude to the existence of $\widetilde{\mathcal{A}}$. Nonetheless, the *a priori* deduction remains an area for exploration.

*Modal characterization of the reduced order Koopman operator*

The acquisition of $\tilde{A}$ by the similarity-matrix expression, or any other forms of $A$ by other methods, signals one's comprehensive possession of a system's underlying dynamics. What can be done to this tangible representation in the form of a matrix is only limited by the bounds of linear algebra. The default procedure for both the DMD and KMD, however, characterises the modal features of $\tilde{A}$ by an eigen decomposition, which is what this work has adopted.

An eigen-Ritz decomposition yields

$$\tilde{A}W = W\Lambda, \quad (AII.10)$$

where $W$ contains the eigenvectors (Ritz vectors) $w_j$, and $\Lambda$ contains the corresponding eigenvalues (Ritz values) $\lambda_j$.

The eigen tuples yield the *exact* DMD modes or the Koopman modes as

$$\Phi = X_2 V \Sigma^{-1} W, \quad (AII.11)$$

where $\Phi$ contains the mode shape $\phi_j$.

Every mode $\phi_j$ corresponds to a physical frequency $\omega_j$ in continuous time,

$$\omega_j = \frac{\Im\{log(\lambda_j)\}}{\Delta t}, \quad (AII.12)$$

and a growth/decay rate $g_j$

$$g_j = \frac{\Re\{log(\lambda_j)\}}{\Delta t}. \quad (AII.13)$$

The sum of the Ritz descriptors expectedly returns the Koopman approximation of the input sequence, or the Koopman system,

$$x_{Koopman,i} = \sum_{j=1}^{r} \phi_j \, exp(\omega_j t_i) \alpha_j, \quad (AII.14)$$

where $\alpha_j$ is the coefficient of weight, or modal amplitude, of $\phi_j$. It is conveniently acquired through a mapping with respect to the initial conditions,

$$\boldsymbol{\alpha} = \boldsymbol{\Phi}^{\dagger} \boldsymbol{x}_1, \quad (AII.15)$$

where the superscript $^{\dagger}$ denotes the Moore-Penrose pseudoinverse.

One may easily assess the accuracy of the Koopman system for each spatial and temporal dimension by comparing the difference between the original and Koopman reconstructed sequences. This work adopts the instantaneous, spatially-averaged, and $l_2$-normalized reconstruction error,

$$\|e\|_{2,i} = \frac{1}{n} \sum_{k=1}^{n} \left[ \left( \frac{x_{Koopman,k,i} - x_{k,i}}{x_{k,i}} \right)^2 \right]^{1/2}, \quad (AII.16)$$

and its rms value

$$\|e\|_{2,rms,\,i} = \left[ \frac{1}{n} \sum_{k=1}^{n} \left( \frac{x_{Koopman,k,i} - x_{k,i}}{x_{k,i}} \right)^2 \right]^{1/2}. \quad (AII.17)$$

as the indices of quantification.

# Appendix III

*Mode Shapes of the $M_1$ of fluctuating, instantaneous, and mean velocity field*

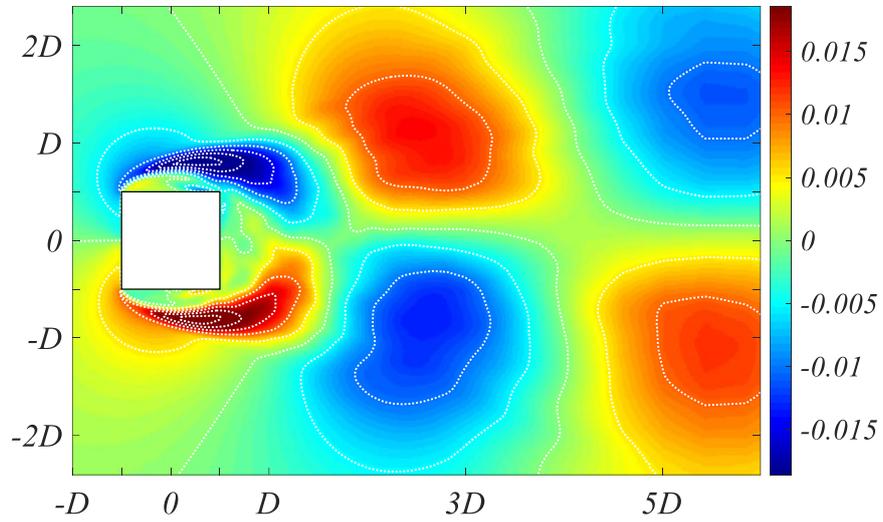

a)

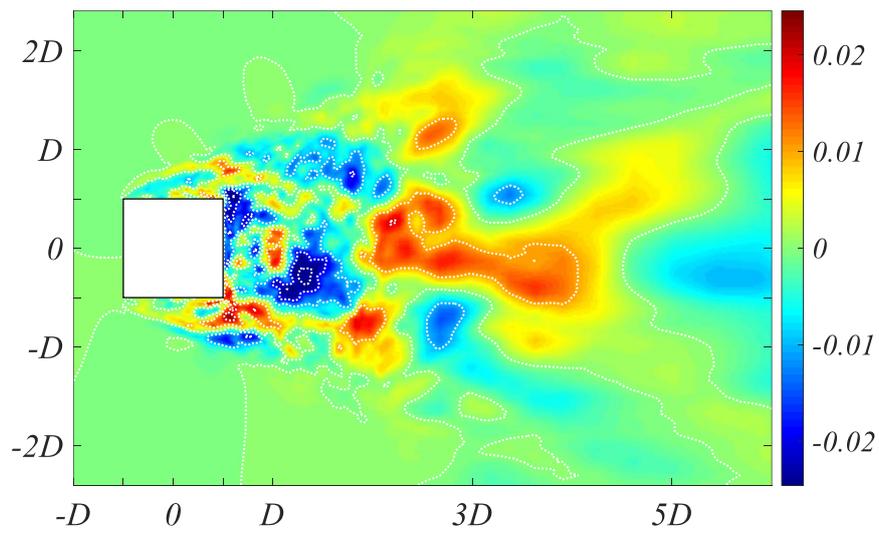

b)

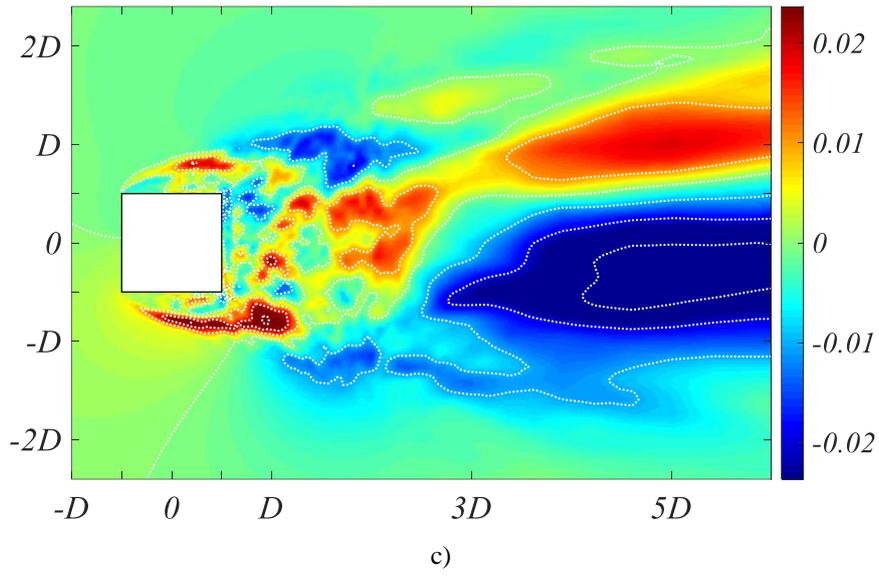

**Fig. AIII.1** Mode shape $\Re(\phi_1)$ of oscillatory DMD modes $M_1$ by sampling 20 oscillation cycles of a) fluctuating velocity $u'$, b) instantaneous velocity $u$, and c) mean velocity $\langle u \rangle$.